\documentclass[twocolumn,showpacs,preprintnumbers,amsmath,amssymb,floatfix,nofootinbib]{revtex4-1}

\usepackage{amsthm}
\usepackage{graphicx}
\usepackage{dcolumn}
\usepackage{bm}

\newcommand{\be}{\begin{equation}}
\newcommand{\ee}{\end{equation}}
\newcommand{\ba}{\begin{eqnarray}}
\newcommand{\ea}{\end{eqnarray}}
\newcommand{\ban}{\begin{eqnarray*}}
\newcommand{\ean}{\end{eqnarray*}}
\newcommand{\non}{\nonumber}
\newcommand{\eq}[1]{(\ref{#1})}
\newcommand{\n}[1]{\label{#1}}
\newcommand{\hh}{\, ,\hspace{0.25cm}}
\newcommand{\hhh}{\, ,\hspace{0.5cm}}

\newcommand{\BM}[1]{{\mbox{\boldmath $#1$}}}

\begin{document}
 
\title{Distorted stationary rotating black holes}
\author{Andrey A. Shoom}
\email{ashoom@phys.ualberta.ca}
\affiliation{Theoretical Physics Institute, University of Alberta, 
Edmonton, AB, Canada,  T6G 2E1}
\begin{abstract}
We study the interior of distorted stationary rotating black holes on the example of a Kerr black hole distorted by external static and axisymmetric mass distribution. We show that there is a duality transformation between the outer and inner horizons of the black hole, which is different from that of an electrically charged static distorted black hole. The duality transformation is directly related to the discrete symmetry of the space-time. The black hole horizons area, surface gravity, and angular momentum satisfy the Smarr formula constructed for both the horizons. We formulate the zeroth, the first, and the second laws of black hole thermodynamics for both the horizons of the black hole and show the correspondence between the local and the global forms of the first law. The Smarr formula and the laws of thermodynamics formulated for both the horizons are related by the duality transformation. The distortion is illustrated on the example of a quadrupole and octupole fields. The distortion fields noticeably affect the proper time of a free fall from the outer to the inner horizon of the black hole along the symmetry semi-axes.  There is some minimal non-zero value of the quadrupole and octupole moments when the time becomes minimal. The minimal proper time indicates the closest approach of the horizons due to the distortion.    
\end{abstract}

\pacs{04.20.Jb, 04.70.-s, 04.70.Bw \hfill  
Alberta-Thy-17-14}

\maketitle

\section{INTRODUCTION}

Electrically neutral astrophysical black holes, stellar or supermassive, located in the centres of most, if not all, galaxies, can (to some extend) be represented by the Kerr black hole solution. However, realistic black holes strongly interact with surrounding matter and fields. Thus, the Kerr black hole solution is highly idealized. However, to consider the complex dynamical processes around a realistic black hole is a formidable problem, which requires advanced numerical computations. In this paper, following the previous works \cite{FS,AFS}, we consider static and axisymmetric type of distortion due to matter located outside a stationary rotating (Kerr) black hole. This type of distortion was considered by Tomimatsu \cite{Tom}, who constructed an implicit form of the exact solution representing a stationary rotating black hole distorted by external static and axisymmetric mass distribution. An explicit from of the solution was found later by Breton, Denisova, Manko, and Garcia \cite{Man,ManQ}. The distortion matter is located at the asymptotic infinity and not included into the solution. As a result, the solution is not asymptotically flat and represents a space-time region in the vicinity of such a black hole. Following Geroch and Hartle who studied static distorted vacuum black holes \cite{GH}, we call such a solution a {\em local black hole}.  

One of the intriguing issues related to the interior of a stationary rotating black hole is stability of its Cauchy horizon. A Cauchy horizon is a null hypersurface which represents a boundary of the region beyond which predictability based on the initial data defined on a spacelike hypersurface outside the black hole breaks down. An analytic continuation of a Kerr black hole beyond its horizons results in an infinite set of other ``universes'' located behind its Cauchy horizon. Whether these new worlds are accessible to an adventurer who is dare to sink behind the black hole horizon depends on the stability of the Cauchy horizon. Such an observer traveling along a timelike world line receives an infinitely blue shifted radiation at the instant of crossing the Cauchy horizon. Penrose argued that this may imply an infinite grow of small perturbations at the Cauchy horizon \cite{Penrose}, what was confirmed in the following works \cite{Per1,Per2,Per3}. There results are not so trivial, however. For example, neglecting radiation backscattered from the Cauchy horizon results in a so-called whimper singularity, which is weak in the sense that the Kretschmann scalar (the trace of the square of the Riemann tensor) is finite. This singularity implies that despite detecting an infinite energy density, a freely falling observer will experience finite tidal forces while crossing the Cauchy horizon \cite{Eli,Kin}. However, the backscattered radiation is present. To analyze its effect, Israel and Poisson  \cite{IP} considered a model of two noninteracting ingoing and outgoing (backscattered) null beams and showed that the presence of the outgoing beam produces a separation between the Cauchy and the inner apparent horizons which results in an infinite growth of the black hole internal mass parameter, the phenomenon which they called {\em mass inflation}, and divergency of the Weyl scalar $\Psi_{2}$ which represents the ``Coulomb'' component of the local curvature. The mass inflation was later analyzed on the example of exact and simplified solution by Ori \cite{Ori}, who showed that despite the tidal forces calculated in the frame of a freely falling observer diverge at the Cauchy horizon, their integral along the observer's world line remains finite. This implies that the mass inflation singularity is not strong enough and it might be possible to cross the Cauchy horizon (for details see \cite{B1,B2,B3,B4,Hamilton}). The following numerical analysis showed instability of the Cauchy horizon \cite{Gne}. However, the subsequent analytical \cite{Bon,Fla} and numerical results \cite{Bra} did not confirm that. In the later work \cite{MarOri}, Marolf and Ori found that in the late-time limit after the perturbation, the inner (Cauchy) horizon ingoing section is replaced by a null, weak curvature singularity (the mass inflation singularity) and the outgoing section of the inner horizon is replaced by a more violent, outgoing shock-wave singularity. The space-time metric is continuous, but not differentiable, across the weak curvature singularity. In the case of the outgoing shock-wave singularity, late-infall-time observers experience a metric discontinuity across the shock, which is at least of the order of unity. 

The purpose of this paper is to consider the interior of a distorted, stationary rotating, and axisymmetric black hole. The distortion is due to external static and axisymmetric distribution of matter. However idealized is the situation when compared to real involved dynamical processes around realistic astrophysical black holes, it still gives important information about the black hole interior, and, in particular, about the effect of distortion on its Cauchy horizon. This is because the external matter distribution notably affects the black hole interior. As it was found in the previous work \cite{AFS}, Cauchy horizon of electrically charged distorted black hole remains regular for such a distribution of matter. This is due to the special duality relations between the metrics on the outer and inner horizon of the black hole. As we shall see, in the case of distorted stationary rotating black hole, there are different duality relation between the black hole horizons. Beside the effect of the distortion on the Cauchy horizon, we shall be interested in thermodynamics of the horizons and study how the distortion affects the black hole interior, in particular, the region between its outer and inner horizons.   

The paper is organized as follows. In Sec. II we present the metric of a distorted stationary rotating black hole. In Sec. III we calculate the area, angular velocity, surface gravity, local mass, and angular momentum and construct Smarr's formula for both the black hole horizons. Form the metric on the black hole horizons we derive a duality transformation between them. In Sec. IV we formulate the zeroth, the first, and the second laws of black hole thermodynamics for both the black hole horizons and find the correspondence between the local and the global forms of the first law. Using the laws of thermodynamics we formulate our morel representing a stationary rotating black hole which is adiabatically distorted by external static and axisymmetric distribution of masses generating gravitational field. In Sec. V we study geometry of the distorted horizons. We calculate Gaussian curvature of the horizon 2-dimensional surfaces and construct their isometric embedding into a flat 3-dimensional space. In Sec. VI we study how the distortion fields affects the proper time of a free fall from the outer to the inner horizon of the black hole along the symmetry semi-axes. In Sec. VII we construct the space-time discrete symmetry from which the duality transformation follows and find a relation between the space-time curvature invariants calculated on the black hole horizons. In addition, we analyze the space-time curvature singularities. We summarize our conclusions in Sec. VIII. 

In this paper we use the following convention of units: $G=c=\hbar=k_{B}=1$, and the sign conventions adopted in \cite{MTW}.

\section{Metric of a distorted Kerr black hole}

Let us begin with the metric representing a {\em local stationary rotating distorted black hole}. An implicit form of such a metric describing a stationary rotating and axisymmetric black hole distorted by external static matter was firstly presented in \cite{Tom}, where the inverse scattering problem technique was applied to derive a stationary axisymmetric solution representing a Kerr black hole in the Weyl background gravitational field of the distorting matter. The solution was given in terms of the Legendre polynomials and differential equations for the corresponding metric functions.  An explicit exact solution of the Einstein equations for the metric functions was derived in \cite{Man} using a representation of the external distorting field in a different form. These results were generalized to the case of distorted Kerr-Newman black hole \cite{ManQ}. Here we present the metric in a slightly different form (with different signs of $\alpha$ and $g_{t\phi}$) than that given in \cite{Man},
\ba\n{II.1}
ds^{2}&=&-\frac{A}{B}e^{2U}dt^{2}+\frac{2}{B}(c_{2}Ae^{2U}-2mC)dtd\phi\non\\
&+&m^{2}c_{1}Be^{2(V-U)}\left(\frac{dx^{2}}{x^{2}-1}+\frac{dy^{2}}{1-y^{2}}\right)\non\\
&+&\frac{e^{-2U}}{AB}\left(m^{2}B^{2}[x^{2}-1][1-y^{2}]\right.\non\\
&-&\left.[c_{2}Ae^{2U}-2mC]^{2}\right)d\phi^{2}\,,
\ea
where $t$ is the time coordinate, $(x,y)$ are the prolate spheroidal coordinates, and $\phi$ is the azimuthal angular coordinate. The metric functions read,
\ba
A&=&(x^{2}-1)(1+fh)^{2}-(1-y^{2})(f-h)^{2}\,,\n{II.2a}\\
B&=&(1+x+fh[x-1])^{2}+([1+y]f+[1-y]h)^{2}\,,\n{II.2b}\\
C&=&(x^{2}-1)(1+fh)(f-h+y[f+h])\non\\
&+&(1-y^{2})(f-h)(1+fh+x[1-fh])\,,\n{II.2c}
\ea
and
\ba
f&=&\alpha\exp\left(2\sum_{n\geq1}a_{n}(x-y)\sum_{l=0}^{n-1}R^{l}P_{l}\right)\,,\n{II.3a}\\
h&=&-\alpha\exp\left(2\sum_{n\geq1}a_{n}(x+y)\sum_{l=0}^{n-1}(-1)^{n-l}R^{l}P_{l}\right)\,,\n{II.3b}
\ea
where
\ba
U&=&\sum_{n\geq0}a_{n}R^{n}P_{n}\,,\n{II.4a}\\
V&=&\sum_{n,k\geq1}\frac{nka_{n}a_{k}}{(n+k)}R^{n+k}(P_{n}P_{k}-P_{n-1}P_{k-1})\,\non\\
&+&\sum_{n\geq1}a_{n}\sum_{l=0}^{n-1}[(-1)^{n-l+1}(x+y)-x+y]R^{l}P_{l}\,.\n{II.4b}
\ea
These functions are given in terms of the Legendre polynomials of the first kind,
\ba
P_{n}&=&P_{n}(xy/R)\hh R=\sqrt{x^{2}+y^{2}-1}\,,\n{II.5a}\\
P_n(-x)&=&(-1)^nP_n(x)\hh P_n(1)=1\,,\n{II.5b}\\
P_{2k}(0)&=&(-1)^{k}\frac{(2k-1)!!}{(2k)!!}\hh P_{2k+1}(0)=0\,,\n{II.5c}\\
k&=&0,1,2,3,...\,,\non\\
(2k-1)!!&=&1\cdot3\cdot5\cdot...\cdot(2k-1)\hh (-1)!!=1\,,\non\\
(2k)!!&=&2\cdot4\cdot6\cdot...\cdot(2k)\hh 0!!=1\,.\non
\ea
The constants $c_{1}$ and $c_{2}$ are expressed in terms of the metric parameters $m$ and $\alpha$ as follows:
\be\n{II.6}
c_{1}=\frac{1}{(1-\alpha^{2})^{2}}\hh c_{2}=\frac{4m\alpha}{1-\alpha^{2}}e^{-2u_{0}}\,,
\ee
where
\be\n{II.7}
u_{0}=\sum_{n\geq0}a_{2n}\,.
\ee
The multipole moments $a_{n}$'s define the distortion.\footnote{These multipole moments sometimes are called the Weyl multipole moments. A relation of the Weyl multipole moments to their relativistic analogues was discussed in \cite{SuenI} for the Schwarzschild black hole distorted by an external field. The general formalism, which includes both the Thorne \cite{Thorne} and the Geroch-Hansen \cite{Ger1,Ger2,Han,Que} relativistic multipole moments is presented in \cite{SuenII}. A relation between the Thorne and the Geroch-Hansen relativistic multipole moments is given in \cite{Beig,Gursel}.} Such distortions result from approximation of a gravitational field in the region far away from its sources. According to the terminology used in the Newtonian gravitational theory and electromagnetism, coefficients in the corresponding multipole expansion of the distortion gravitational field are called {\em interior multipole moments}. The distortion fields $U$ and $V$ defined by the interior multipole moments are regular and smooth at the black hole horizons.\footnote{Note that despite $U$ satisfies the Laplace equation, it is a relativistic field. In order to construct the corresponding Newtonian analogue of the field, one has to take the nonrelativistic limit $\lim_{c^{2}\to\infty}c^{2}U(x,y,c^{2})$, where $c$ is the speed of light (see, e.g., \cite{Eh,Qu}).} The {\em exterior multipole moments} correspond to the sources located inside the black hole and the corresponding distortion field is given in terms of the Legendre polynomials of the second kind \cite{ManQ}. According to the Kerr black hole uniqueness theorem (see, e.g., \cite{Carter,Carter2,ChrCo}), the Kerr black hole is the only stationary, asymptotically flat, vacuum black hole with regular horizons. Thus, such sources make the black hole horizons singular \cite{MN,ManQ}. 

The dominant term in the expansion of the distortion field is the monopole and in our case it represents background distortion defined by a monopole moment $a_{0}$. The next term is the dipole defined by a dipole moment $a_{1}$, which according to the black hole equilibrium condition (see Eq.\eq{II.8} below), is related to the higher order multipole moments.  The next term is the quadrupole, which is defined by a quadrupole moment $a_{2}$, and which is followed by the octupole defined by an octupole moment $a_{3}$. Here we shall consider only these subleading terms in the multipole expansion of the distortion field.

To have the horizons free of conical singularities on the symmetry axis $y=\pm1$ the multipole moments have to satisfy the following condition:
\be\n{II.8}
\sum_{n\geq0}a_{2n+1}=0\,.
\ee
This condition is sometimes called the black hole equilibrium condition \cite{Chandrabook}.

According to the way the solution is constructed (see \cite{Tom,Man,ManQ}), the sources of the distorting field, regardless the generated solution represents non-stationary rotating (static) or stationary rotating (stationary) distorted black hole, satisfy the strong energy condition (see, e.g. \cite{GH}),
\be\n{8a}
U\leq0\,,
\ee
which implies 
\be\n{8b}
u_{0}\leq0\,,
\ee
The expressions above define a {\em local stationary rotating distorted black hole}. The solution represents a distorted non-extremal Kerr black hole with $|\alpha|\in(0,1)$. Without the loss of generality, we shall consider nonnegative values of $\alpha$. The case $\alpha=0$ corresponds to a distorted static (Schwarzschild) black hole. We shall study properties of the {\em local distorted stationary rotating black hole} in the following sections.

Using the coordinate transformations 
\be\n{II.9}
x=\frac{r}{m}-\frac{1+\alpha^{2}}{1-\alpha^{2}}\hhh y=\cos\theta\,,
\ee
and removing the distortion by making all the multipole moments $a_{n}$'s vanish we derive the Kerr metric solution given in the Boyer-Lindquist coordinates (see, e.g., \cite{FN}),
\ba\n{II.10}
ds^2&=&-\left(1-\frac{2Mr}{\Sigma}\right)dt^2-\frac{4Mar\sin^2\theta}{\Sigma}dtd\phi\non\\
&+&\frac{\Sigma}{\Delta}\,dr^2+\Sigma\,d\theta^2\\
&+&\left(r^2+a^2+\frac{2Ma^2r\sin^2\theta}{\Sigma}\right)\sin^2\theta d\phi^2\,,\non\\
\Sigma&=&r^2+a^2\cos^2\theta\hhh \Delta=r^2-2Mr+a^2\,,\non
\ea 
where the mass $M$ and the angular momentum per unit mass, $a=J/M$, of an undistorted (Kerr) black hole are related to the parameters $\alpha$ and $m$ as follows:
\be\n{II.10a}
M=m\left(\frac{1+\alpha^{2}}{1-\alpha^{2}}\right)\hhh a=\frac{2m\alpha}{1-\alpha^{2}}\,.
\ee
The quantities $M$ and $J$ are defined as 
\ba
M&=&-\frac{1}{4\pi}\oint_{S_{\infty}} d^{2}\Sigma_{\mu\nu}\nabla^\mu\xi^\nu_{(t)}\,,\n{II.11a}\\
J&=&\frac{1}{8\pi}\oint_{S_{\infty}}d^{2}\Sigma_{\mu\nu}\nabla^\mu\xi^\nu_{(\phi)}\,,\n{II.11b}
\ea
where $\xi^\mu_{(t)}=\delta^\mu_t$ is a timelike Killing vector normalized at the asymptotic infinity, $\BM{\xi}^2_{(t)}=-1$, $\xi^\mu_{(\phi)}=\delta^\mu_\phi$ is a space-like rotational Killing vector, and
\be\n{II.12}
d^{2}\Sigma_{\mu\nu}=\frac{1}{2!}\sqrt{-g}\,\varepsilon_{\mu\nu\lambda\sigma}\,
dx^{\lambda}\wedge dx^{\sigma}\hhh \varepsilon_{tr\theta\phi}=+1\,,
\ee
is the area element of a two-dimensional closed space-like surface at the spatial infinity, $S_{\infty}$. The mass \eq{II.11a} and the angular momentum \eq{II.11b}  can be presented as follows (see, e.g., \cite{FN,BCH}):
\ba
M&=&\int_{\Sigma}(2T^{\mu}_{\nu}-T^{\alpha}_{\alpha}\delta^{\mu}_{\nu})\xi^\nu_{(t)}d\Sigma_{\mu}+M_{H}\,,\n{II.12a}\\
J&=&\int_{\Sigma}T^{\mu}_{\nu}\xi^\nu_{(\phi)}d\Sigma_{\mu}+J_{H}\,,\n{II.12b}
\ea
where $T^{\mu}_{\nu}$ is the energy-momentum tensor which defines matter and fields outside the black hole, $M_{H}$ and $J_{H}$ are the local mass and angular momentum of the black hole alone, which are defined by the integrals \eq{II.11a} and \eq{II.11b} over the two-dimensional horizon surface $S_{H}$. In the absence of the external matter and fields, i.e., in the case of the Kerr black hole, we have
\be\n{II.12c}
M=M_{H}\hh J=J_{H}\,.
\ee
In what follows, we shall drop the subscript $H$. 

\section{Horizons, Smarr's formula, and the duality transformation}

In this section we define the outer and inner horizons of a distorted stationary rotating black hole, calculate area of the horizons 2-dimensional spacelike surface, angular velocity, surface gravity, the local mass, and angular momentum, which allow to construct Smarr's formula for both the horizons. We construct metrics on the horizons and find a duality transformation connecting these metrics.  

\subsection{Horizons and Smarr's formula}

The black hole horizons, which are Killing horizons of the space-time, are at 
\be\n{III.1}
x=\pm1\,,
\ee
where the upper sign stands for the event (outer) horizon and the lower sign stands for the Cauchy (inner) horizon. In what follows (except for Sec. VI), to indicate that a quantity (...) is calculated at the black hole horizons we shall use the subscripts $\pm$ and denote such a quantity as $(...)_{\pm}$. Accordingly, the upper sign stands for the outer horizon and the lower one stands for the inner horizon.

The horizon surface areas can be calculated from the metric \eq{II.1} by taking $t=const$ and using the property of the Legendre polynomials \eq{II.5b}. Substituting \eq{III.1} into the metric functions we derive the following useful relations:
\ba
f_{\pm}&=&\alpha\,e^{\mp2u_{\pm}(y)}\hh h_{\pm}=-f_{\pm}\,,\n{III.2a}\\
U_{\pm}&=&u_{\pm}(y)+u_{0}\hh V_{\pm}=\pm2u_{\pm}(y)\,,\n{III.2b}
\ea
where
\be\n{III.3}
u_{\pm}(y)=\sum_{n\geq0}(\pm1)^{n}a_{n}y^{n}-u_{0}\,.
\ee
The function $u_{\pm}(y)$ has the symmetry properties
\be\n{III.3a}
u_{\pm}(y;a_{2n+1})=u_{\pm}(-y;-a_{2n+1})\,.
\ee

The horizon surface areas read
\be\n{III.4}
{\cal A}_{\pm}=16\pi m^{2}e^{-2u_{0}}\frac{(1+\alpha^{\pm2})}{(1-\alpha^{\pm2})^{2}}\,.
\ee
The angular velocity of the black hole horizons is defined as follows:
\be\n{III.7}
\Omega_{\pm}=-\frac{g_{t\phi}}{g_{\phi\phi}}\rvert_{x=\pm1}\,,
\ee
and we have
\be\n{III.8}
\Omega_{\pm}=\frac{\alpha^{\pm1}(1-\alpha^{2})}{2m(1+\alpha^{2})}e^{2u_{0}}\,.
\ee
The surface gravity is defined as follows:
\be\n{III.5}
\kappa^{2}_{\pm}=-\frac{1}{2}(\nabla^{\alpha}\chi^{\beta}_{\pm})(\nabla_{\alpha}\chi_{\pm\beta})|_{x=\pm1}\,,
\ee
where 
\be\n{III.5a}
\BM{\chi}_{\pm}=\BM{\xi}_{(t)}+\Omega_{\pm}\BM{\xi}_{(\phi)}\,.
\ee
For the black hole horizons we have
\be\n{III.6}
\kappa_{\pm}=\frac{(1-\alpha^{\pm2})^{2}}{4m(1+\alpha^{\pm2})}e^{2u_{0}}\,.
\ee
The local values of the mass and angular momentum are calculated from \eq{II.11a} and \eq{II.11b}, respectively, where integration is preformed over the horizon surface $S_{H}$. We derive
\ba
M&=&m\left(\frac{1+\alpha^{2}}{1-\alpha^{2}}\right)\,,\n{III.9a}\\
J&=&2m^{2}e^{-2u_{0}}\frac{\alpha(1+\alpha^{2})}{(1-\alpha^{2})^{2}}\,.\n{III.9b}
\ea
The area, angular velocity, surface gravity, the local mass, and angular momentum satisfy Smarr's formula,
\be\n{III.10}
M=\pm\frac{1}{4\pi}\kappa_{\pm}{\cal A}_{\pm}+2\Omega_{\pm}J\,,
\ee
and the horizon areas satisfy the area relations \cite{AH1,AH2,AH3}, 
\be\n{III.11}
{\cal A}_{-}<\sqrt{{\cal A}_{-}{\cal A}_{+}}<{\cal A}_{+}\hh \sqrt{{\cal A}_{-}{\cal A}_{+}}=8\pi J\,.
\ee 

\subsection{Duality transformation}

In the previous works \cite{FS,AFS}, a duality transformation between the horizon and the {\em stretched} singularity of a Schwarzschild distorted black hole and between the outer and inner horizons of a distorted  electrically charged (Reissner-Nordstr\"om) black hole was established. An analogous duality transformation exists for 5-dimensional generalizations of these distorted black holes \cite{ASP,AS}. Here we shall establish a duality transformation between the outer and inner horizons of a distorted stationary rotating black hole. As we shall see, the duality transformation is different from that of the static distorted black holes.  

To establish a duality transformation we consider the metrics on the horizons of the black hole. Using the metric \eq{II.1} we derive the metrics on the horizons,
\ba
d\sigma_{\pm}^{2}&=&F_{\pm}(y)\left(\frac{\alpha^{\pm2}e^{2u_{0}}}{1+\alpha^{\pm2}}dt^{2}\mp
\frac{4m\alpha^{\pm1}}{1-\alpha^{\pm2}}dtd\phi\right)\,\non\\
&+&\frac{{\cal A}_{\pm}}{4\pi}\left(\frac{dy^{2}}{F_{\pm}(y)}+F_{\pm}(y)d\phi^{2}\right)\,.\n{III.12}
\ea
Here
\be\n{III.13}
F_{\pm}(y)=\frac{(1+\alpha^{\pm2})(1-y^{2})}{\alpha^{\pm2}y^{2}+e^{4u_{\pm}(y)}}e^{2u_{\pm}(y)}\,.
\ee
Using these expressions we can construct the {\em duality transformation} which relates the metrics on the horizons:
\be\n{III.14}
u_{\pm}\to u_{\mp}\hh \alpha\to \alpha^{-1}\,.
\ee
This transformation corresponds to a ``switch'' between the outer and the inner horizon of a distorted stationary rotating black hole. Using the expression \eq{III.3} we can present the duality transformation in terms of the Weyl multipole moments, 
\be\n{III.15}
a_{2n}\to a_{2n}\hh a_{2n+1}\to -a_{2n+1}\hh \alpha\to \alpha^{-1}\,.
\ee
Alternatively, one can present this transformation as an exchange between the symmetry semi-axes $y=-1$, $y=1$, while keeping the values of the multipole moments unchanged,
\be\n{III.16}
y\to -y\hh \alpha\to \alpha^{-1}\,.
\ee 
Note that this duality transformation is different from the duality transformation between the outer and the inner horizon of an electrically charged distorted black hole \cite{AFS}, or the duality transformation between the horizon and the {\em stretched} singularity of a distorted Schwarzschild black hole \cite{FS}, where it corresponds to reverse of signs of even multipole moments. Let us note as well that an explicit relation for the metric on the inner horizon in terms of that on the outer horizon was derived for the Ernst potential in the works by Ansorg and Hennig \cite{AH1,AH2,AH3}, without having an explicit solution. It allowed the authors to construct the area relations \eq{III.11} for axisymmetric and stationary distorted black holes.

The expressions for the horizon surface areas \eq{III.4}, surface gravity \eq{III.6}, angular velocity \eq{III.8}, local mass \eq{III.9a}, and angular momentum \eq{III.9b} are related through the duality transformation as follows:
\ba\n{III.17}
&&{\cal A}_{\pm}\to{\cal A}_{\mp}\hh \kappa_{\pm}\to \kappa_{\mp}\hh \Omega_{\pm}\to-\Omega_{\mp}\,,\non\\
&&M\to-M\hh J\to J\,.
\ea 
These relations transform Smarr's formula \eq{III.10} for the outer and inner horizons into each other.

\section{Thermodynamics of the distorted black Hole}

In this Section we present the laws of thermodynamics for a distorted stationary rotating black hole. The laws of thermodynamics for a static distorted black hole, Schwarzschild and Reissner-Nordstr\"om ones, were presented by Geroch and Hartle \cite{GH} and by Fairhurst and Krishnan \cite{FK}, respectively. A black hole thermodynamic variables, such as energy, entropy, temperature, etc. are related to the black hole ``mechanical'' variables (see \cite{BCH}), such as mass, horizon surface area, surface gravity, etc., respectively. Such a relation was originally conjectured by Bekenstein \cite{Bek} and established by Hawking \cite{Haw}.    

\subsection{The zeroth law}  

According to the zeroth law of black hole thermodynamics, black hole temperature is constant over the black hole horizon. The temperature is defined in terms of surface gravity as 
\be\n{IV.1}
T_{\pm}=\frac{\kappa_{\pm}}{2\pi}\,.
\ee
The surface gravity is defined up to an arbitrary constant which depends on the normalization of the timelike Killing vector $\BM{\xi}_{(t)}$. This definition requires a proper normalization of the Killing vector at the spatial infinity. As far as only a normalization constant is involved, the normalization does not affect the zeroth law. 

One can see from the expression \eq{III.6}, that the zeroth law holds for both the horizons of the distorted stationary rotating black hole. The temperature $T_{+}$ is associated with the black hole outer horizon, while the temperature $T_{-}$ is associated with the black hole inner (Cauchy) horizon. One may think that $T_{-}$ is rather a dubious thermodynamic variable. However, according to the description of black holes within string theory \cite{Hor}, the outer horizon thermodynamics is considered as the sum of the thermodynamics corresponding to the left- and right-moving string excitations. According to such a duplicate nature of the horizon thermodynamics, one can view the inner horizon thermodynamics as the difference of the thermodynamics corresponding to the right- and left-moving excitations of the string \cite{Lar,CvetLar,CvetLar2}. In this picture, the thermodynamic variables corresponding to the outer horizon are mapped to the thermodynamic variables corresponding to the inner horizon, and vice versa. In our case, such a map is represented by the duality transformation, Eqs.\eq{III.15}--\eq{III.17}. In what follows, we shall adopt this picture of thermodynamics of the distorted black hole and consider thermodynamics of both the horizons. 

\subsection{The first law}

The first law of the black hole thermodynamics defines a relation between two nearby equilibrium states of a thermodynamic system which includes a black hole. These states are related by a change in the system's energy, entropy, and other parameters, e.g., angular momentum. There are different forms of the first law, which are defined according to the system in question. Here we shall consider global and local forms of the first law.  

\subsubsection{The global form of the first law}

The global form of the first law takes into account the effects of the external matter on the black hole. The solution \eq{II.1} represents a {\em local black hole} for it doesn't include the external matter, which distorts it. Assuming that the solution can be analytically extended to achieve asymptotic flatness, one can include sources of the external matter into the solution. An analytic extension is achieved by requiring that the distortion fields $U$ and $V$ vanish at the asymptotic infinity and by extending the space-time manifold \cite{GH}. Due to energy-momentum tensor of the sources, the extended solution will not satisfy the vacuum Einstein equations in the region where the external matter is located. As a result of the extension, we have a vacuum region containing the {\em local black hole}, the external to it region containing the sources, and the asymptotically flat region outside the sources. The extension allows to normalize the timelike Killing vector $\BM{\xi}_{(t)}$ at the spatial infinity as $\BM{\xi}_{(t)}^2=-1$. Then, the Komar mass $M$ of the distorted stationary rotating black hole (without the external matter) \eq{II.11a} equals to its local mass [cf. Eq.\eq{II.12a}].

To derive the global form of the first law one can use the expression for the black hole horizon area \eq{III.4} and the definition of the black hole entropy,
\be\n{IV.2}
S_\pm=\frac{\mathcal{A}_\pm}{4}\,.
\ee
Taking the differential of \eq{III.4} and using the expressions for the surface gravity \eq{III.6}, angular velocity \eq{III.8}, local mass \eq{III.9a}, and angular momentum \eq{III.9b} we derive the global form of the first law, 
\be\n{IV.3}
\delta M=\pm T_\pm\delta S_\pm + \Omega_{\pm}\delta J+M\delta u_{0}\,.
\ee       
The term $M\delta u_{0}$ represents the work done on the black hole by the variation of the external potential $u_{0}$ due to the distorting matter. If the distortion field changes adiabatically, i.e. if $\delta S_+=0$, such that neither matter nor gravitational waves cross the black hole outer horizon \cite{GH}, and in addition, the black hole angular momentum $J$ does not change, which implies that $\delta\alpha=0$, so that $\delta S_{-}=0$, then the work $M\delta u_{0}$ results in the change of the black hole mass $\delta M$, i.e.,
\be\n{IV.3a}
\delta M=M\delta u_{0}\,.
\ee
Integrating this relation we derive
\be\n{IV.3b}
M=M_{0}e^{u_{0}}\,,
\ee
where $M_{0}$ is the local mass in the absence of the distortion. The factor $e^{u_{0}}$ can be considered as a redshift factor due to the distortion.
  
\subsubsection{The local form of the first law}

The local form of the first law doesn't take into account the effects of the external matter on the black hole. It can be defined by observers who live near the black hole and attribute the local gravitational field to the black hole alone. These observers consider the black hole as an isolated, undistorted object.  They assume that there is no other matter present and the space-time is asymptotically flat. Such observers define the outer horizon area $\mathcal{\tilde A}_+$, the angular velocity $\tilde{\Omega}_+$, surface gravity $\tilde{\kappa}_+$, the black hole local mass $\tilde{M}$ and the angular momentum $\tilde{J}$, such that they satisfy Smarr's formula for the Kerr black hole,
\be\n{IV.4a}
\tilde{M}=\frac{1}{4\pi}\tilde{\kappa}_{+}\tilde{\cal A}_{+}+2\,\tilde{\Omega}_{+}\tilde{J}\,.
\ee
We shall use the sign $\,\tilde{}\,\,$ for the quantities defined by the local observers. According to the observers measurement of the local quantities, one has
\be\n{IV.5}
{\cal A}_{+}=\tilde{\cal A}_{+}\hh J=\tilde{J}\,.
\ee
These relations together with \eq{III.4} and \eq{III.9b} give
\be
\alpha=\tilde{\alpha}\hh m=\tilde{m}\,e^{u_{0}}\,.\n{IV.6}
\ee
Smarr's formula for the black hole inner horizon reads [cf. Eq.\eq{III.10}]
\be\n{IV.4b}
\tilde{M}=-\frac{1}{4\pi}\tilde{\kappa}_{-}\tilde{\cal A}_{-}+2\,\tilde{\Omega}_{-}\tilde{J}\,.
\ee
Then, the expressions \eq{III.4}, \eq{III.6}, \eq{III.8}, \eq{III.9a}, and \eq{III.9b} give
\ba\n{IV.7}
{\cal A}_{-}&=&\tilde{\cal A}_{-}\hh \kappa_{\pm}=\tilde{\kappa}_{\pm}e^{u_{0}}\hh \Omega_{\pm}=\tilde{\Omega}_{\pm}e^{u_{0}}\,,\non\\
M&=&\tilde{M}e^{u_{0}}\hh J=\tilde{J}\,.
\ea
Accordingly, the local form of the first law of black hole thermodynamics reads
\be\n{IV.9}
\delta\tilde{M}=\pm\tilde{T}_\pm\delta \tilde{S}_\pm+\tilde{\Omega}_\pm\delta\tilde{J}\,.
\ee 
The relations \eq{IV.5}, \eq{IV.6}, and \eq{IV.7} establish the correspondence between the local and the global forms of the first law. A comparison of the relation \eq{IV.3b} and the relation between $M$ and $\tilde{M}$ given in the expression \eq{IV.7} justifies that the local mass, as it is measured by the observers, corresponds to undistorted stationary rotating black hole.   

\subsection{The second law and the model}

The second law of black hole thermodynamics implies that if a thermodynamic system consists of a black hole alone, its entropy, thus the outer (event) horizon area, classically never decreases,
\be\n{IV.10}
\delta S_{+}\geq0\iff\delta{\cal A}_{+}\geq0\,.
\ee
Let us discuss a model of a stationary rotating black hole distorted by external, static, and axisymmetric distribution of masses which generate a gravitational field. This distribution is defined by the Weyl multipole moments $a_{n}$'s. A change in the matter distribution causes change in their values. For example, we can bring the masses closer to the black hole or move them further from it. Let us now consider an idealistic situation, such that the change is infinitesimally slow, so that no gravitational waves, which may enter the black hole, are generated (see \cite{GH}). In this case, the distortion is adiabatic, and, as a result, the black hole outer horizon area, which is a measure of its entropy, remains constant. Assume now that at the beginning there is no distortion, i.e., all $a_{n}$'s vanish, and let the corresponding parameters of the solution \eq{II.1} be $\alpha'$ and $m'$. Then, the local quantities such as ${\cal  A}'_{+}$ and $J'$ can be expressed in terms of $\alpha'$ and $m'$ as follows [cf. \eq{III.4} and \eq{III.9b}]:
\be\n{IV.11}
{\cal A}'_{+}=16m'^{2}\frac{(1+\alpha'^{2})}{(1-\alpha'^{2})^{2}}\hh J'=2m'^{2}\frac{\alpha'(1+\alpha'^{2})}{(1-\alpha'^{2})}\,.
\ee
An adiabatic change (which preserves the axial symmetry) in the values of $a_{n}$'s doesn't change ${\cal A}_{+}$ and $J$ [cf. \eq{IV.5}]. During the adiabatic change of the distortion field they remain equal to those corresponding to an undistorted stationary rotating black hole,
\be\n{IV.12a}
{\cal A}_{+}={\cal A}'_{+}\hh J=J'\,.
\ee
These equalities imply the following relation between the parameters:
\be
\alpha=\alpha'\hh m=m'\,e^{u_{0}}\,.\n{IV.12b}
\ee
This relation is identical to the relation \eq{IV.6}, which implies that in the case of adiabatic distortion the local observers discussed in the previous subsection detect the actual values of the black hole area and angular momentum, regardless that they completely ignore the distortion effects. In what follows, to study stationary rotating distorted black holes we shall consider this model and use these relations.     

\begin{figure*}[htb]
\begin{center}
\hspace{0cm}
\ba
&&\hspace{0cm}\includegraphics[width=7.0cm]{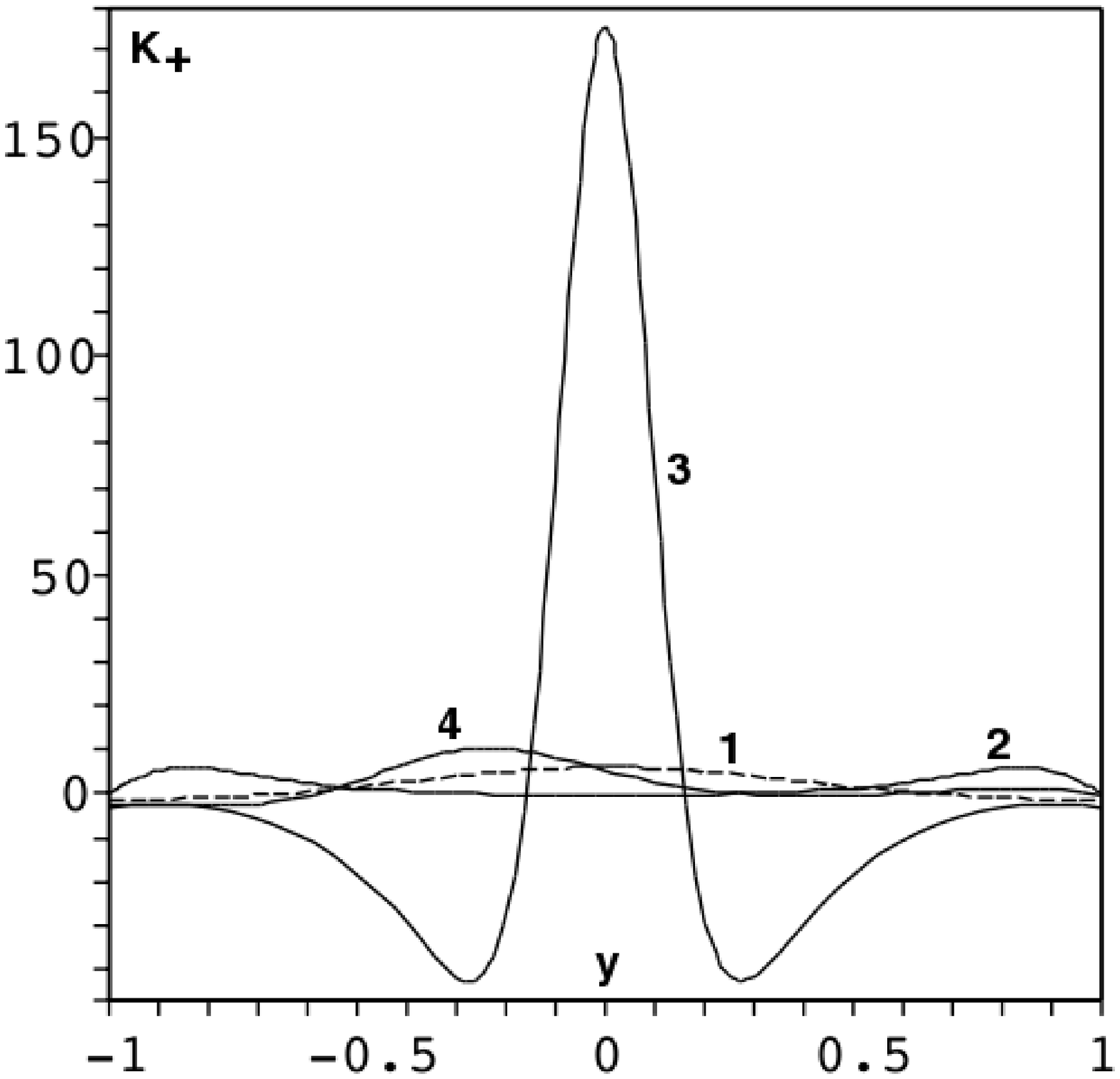}
\hspace{1.5cm}\includegraphics[width=7.0cm]{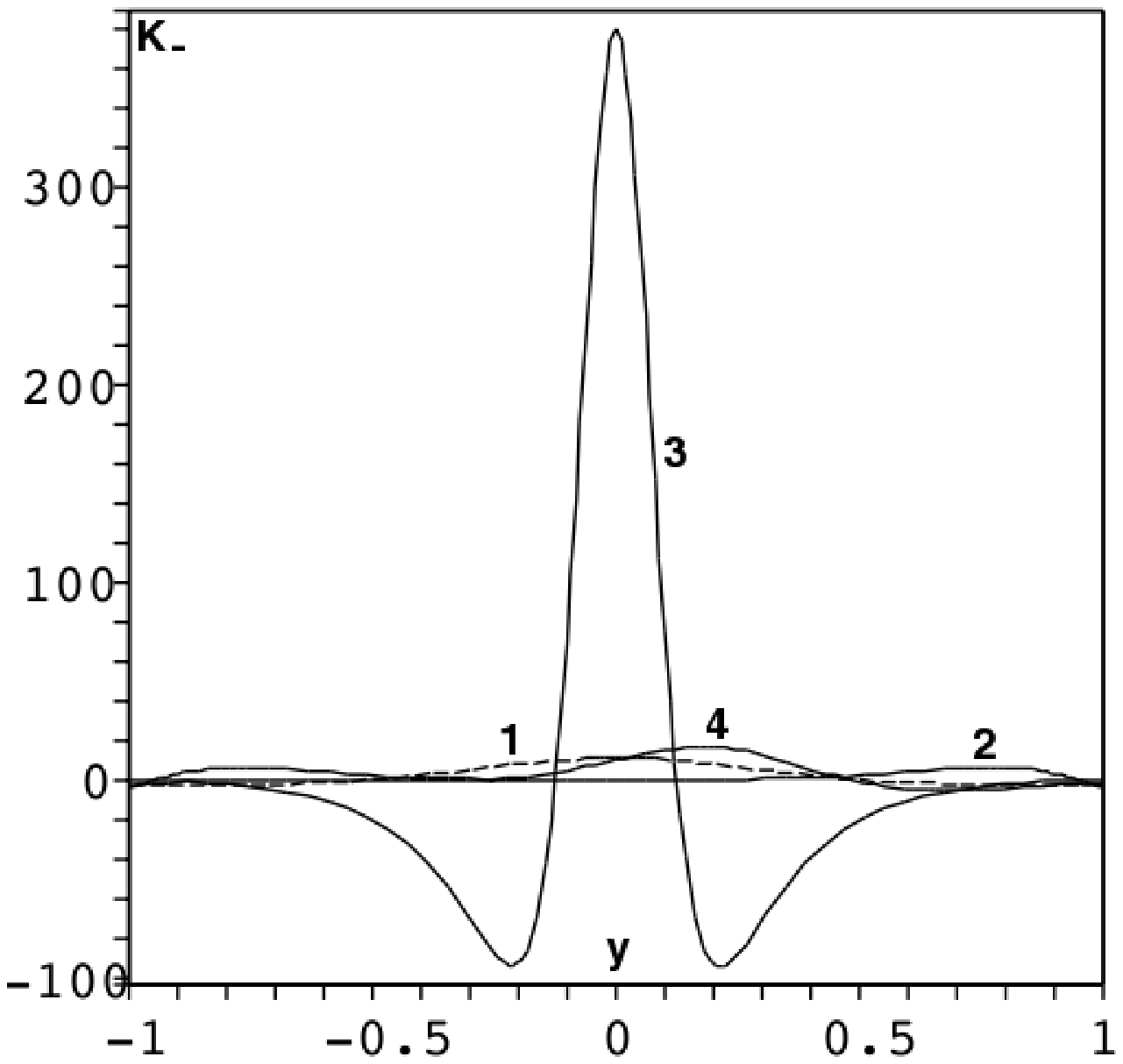}\non\\
&&\hspace{3.5cm}({\bf a})\hspace{8.0cm}({\bf b})\non\\
&&\hspace{0cm}\includegraphics[width=7.0cm]{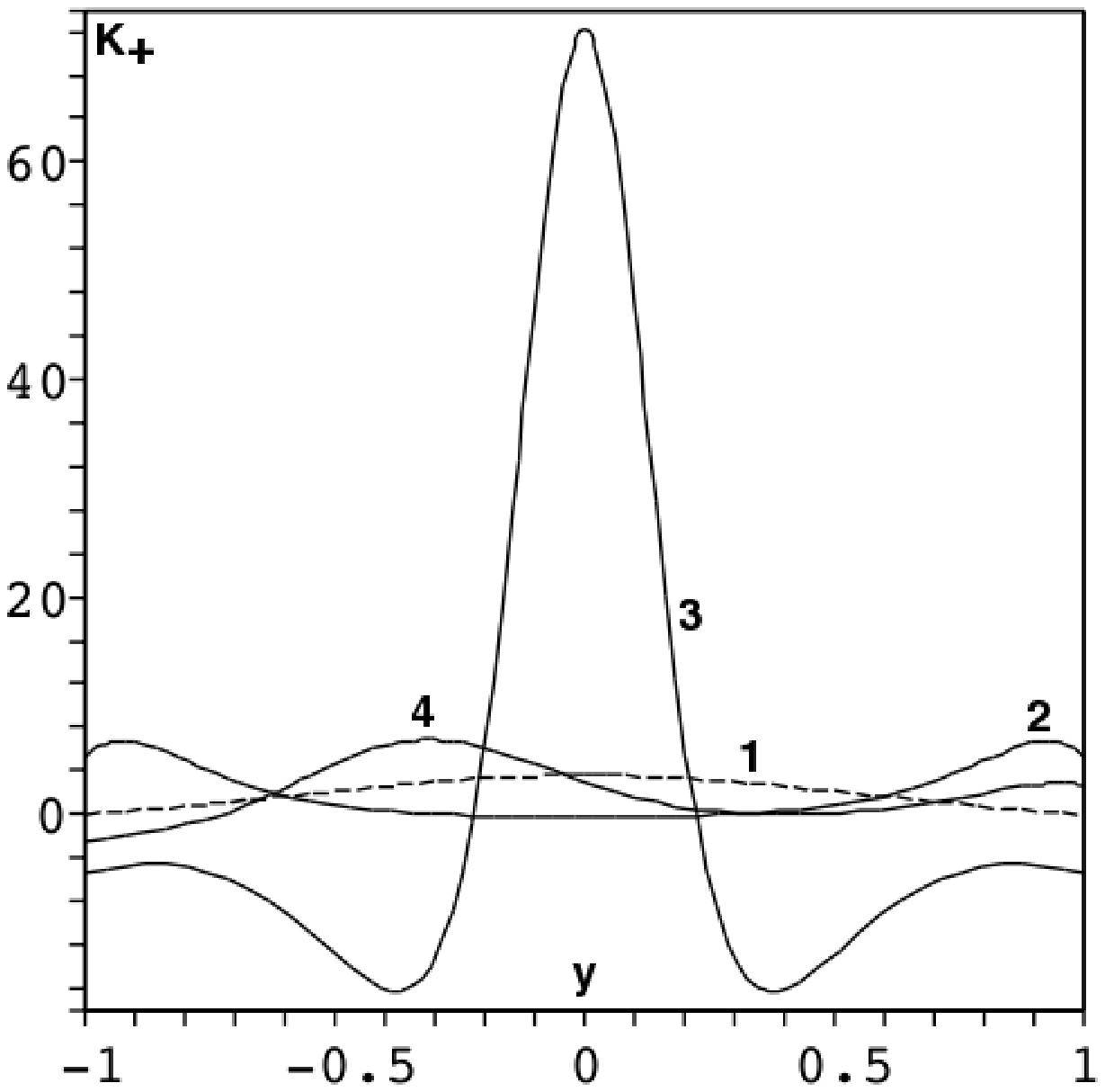}
\hspace{1.5cm}\includegraphics[width=7.0cm]{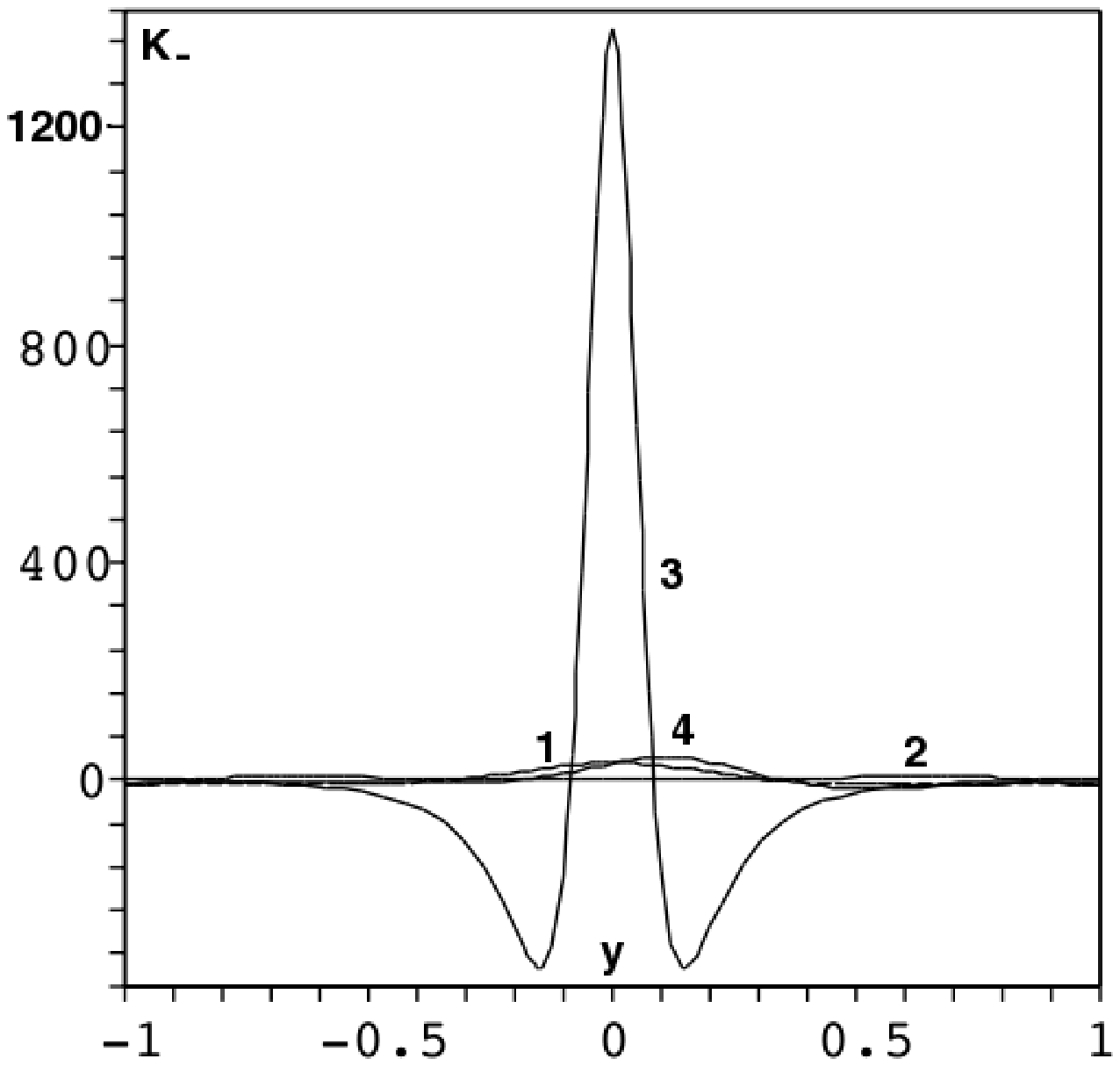}\non\\
&&\hspace{3.5cm}({\bf c})\hspace{8.0cm}({\bf d})\non\\
&&\hspace{0cm}\includegraphics[width=7.0cm]{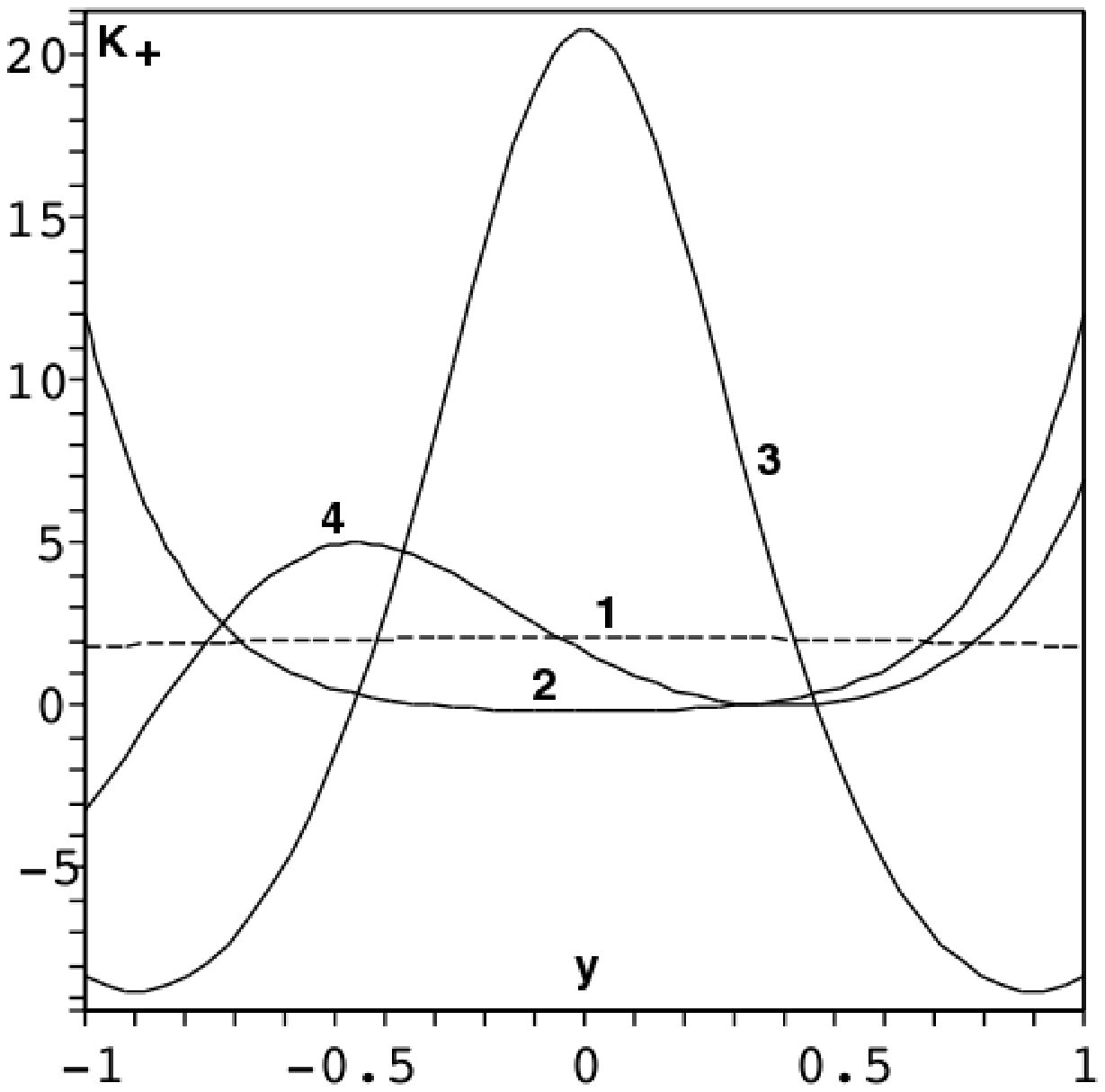}
\hspace{1.5cm}\includegraphics[width=7.0cm]{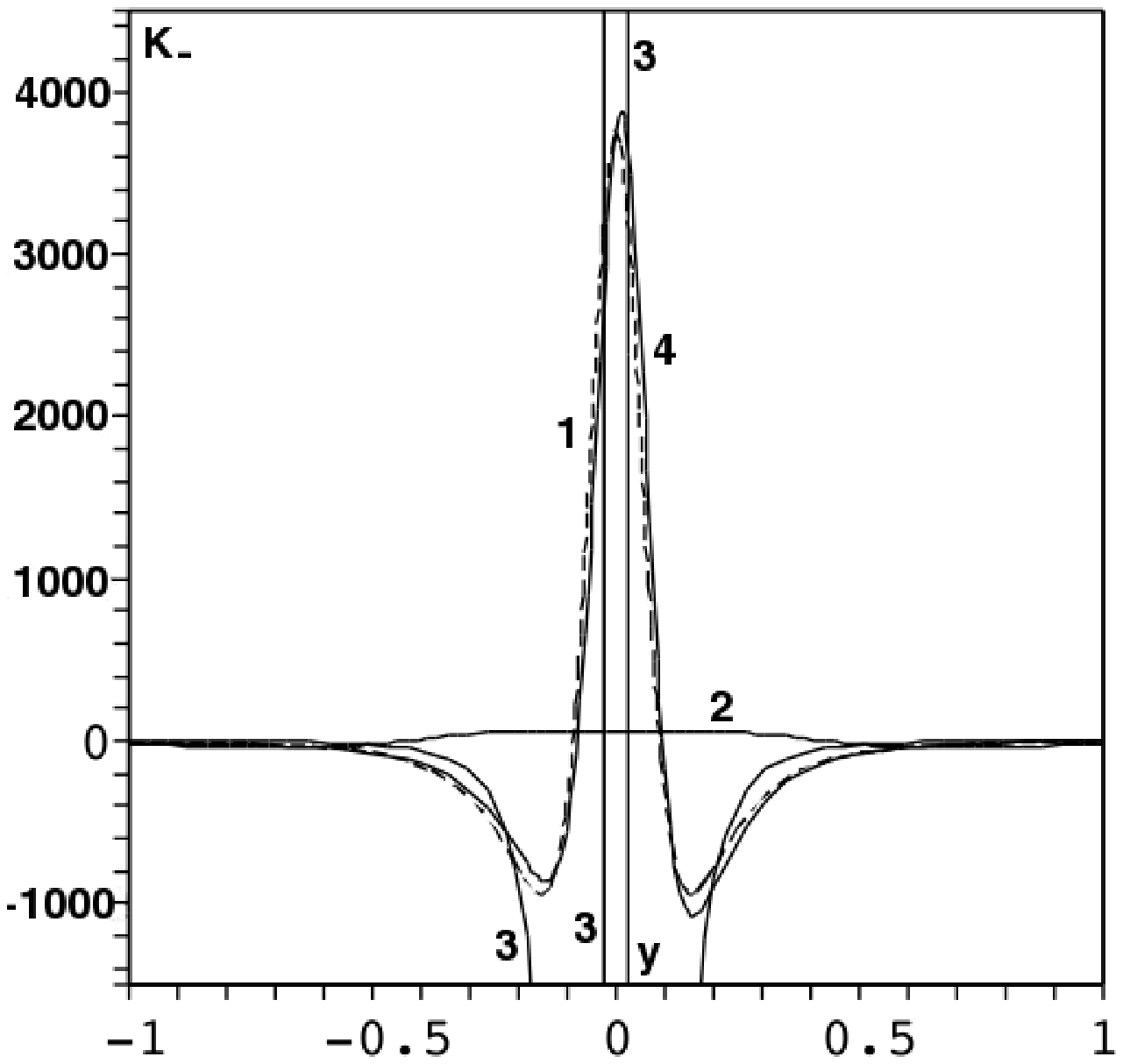}\non\\
&&\hspace{3.5cm}({\bf e})\hspace{8.0cm}({\bf f})\non
\ea
\caption{Gaussian curvature of the horizon surfaces. Dashed line 1 is for undistorted horizons, line 2 is for the quadrupole distortion of $a_{2}=-2/3$, line 3 is for the quadrupole distortion of $a_{2}=2/3$, and line 4 is for the octupole distortion of $a_{3}=-1/3$. Plots (a) and (b) are for $a_{*}=0.99$, $\alpha\approx0.868$. Plots (c) and (d) are for $a_{*}=\sqrt{3}/2\approx0.866$, $\alpha=1/\sqrt{3}\approx0.577$. Plots (e) and (f) are for $a_{*}=0.3$, $\alpha\approx0.154$. To make the illustration better, we cut out some portions of the line 3 in plot (f) near the maximal value $K_{-}\approx2.02\times10^{5}$ at $y=0$ and the minimal ones $K_{-}\approx-5.05\times10^{4}$ at $y\approx\pm0.04$.} \label{f1} 
\end{center}
\end{figure*}
\begin{figure*}[htb]
\begin{center}
\hspace{0cm}
\ba
&&\hspace{0cm}\includegraphics[width=7.0cm]{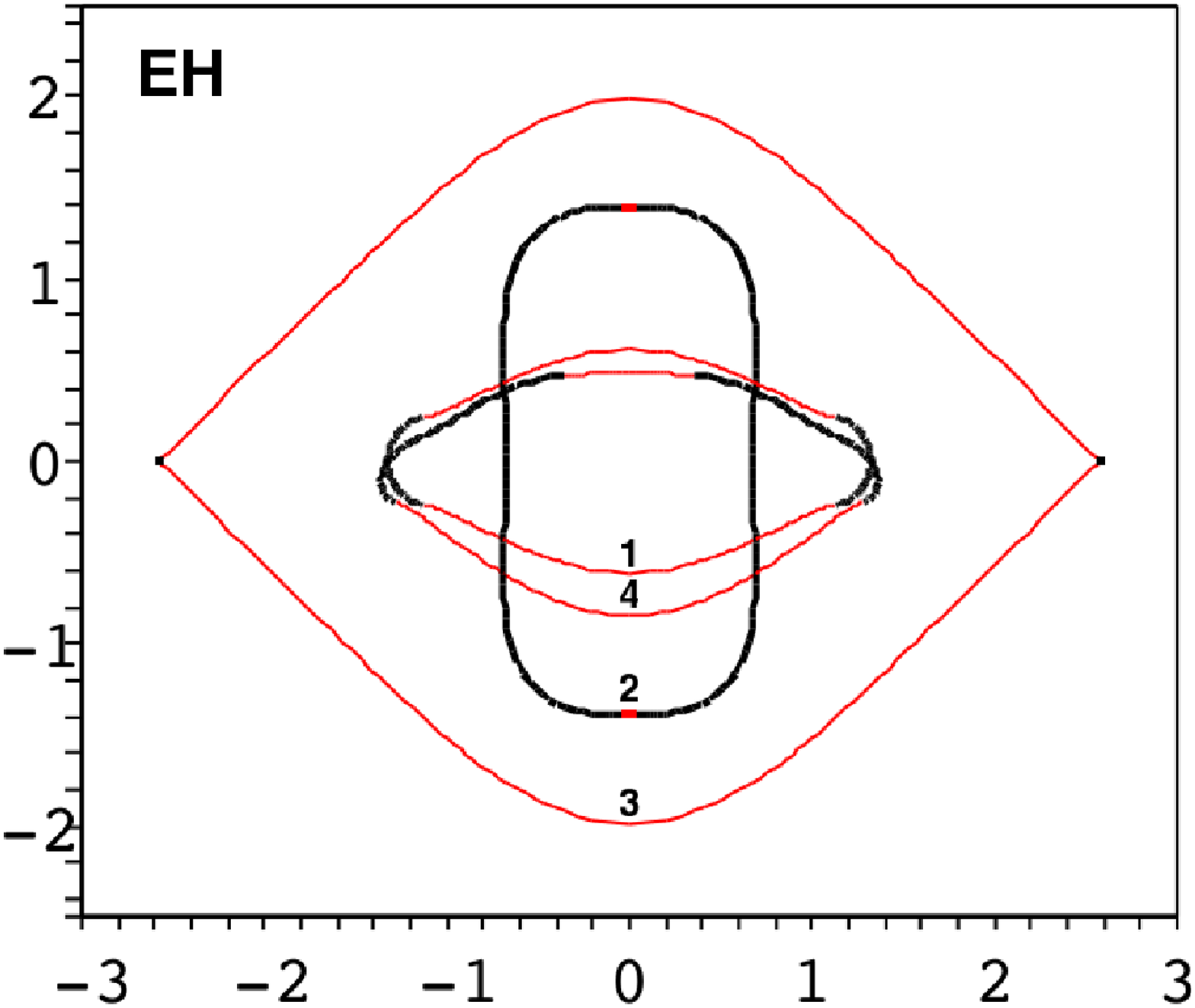}
\hspace{1.5cm}\includegraphics[width=7.0cm]{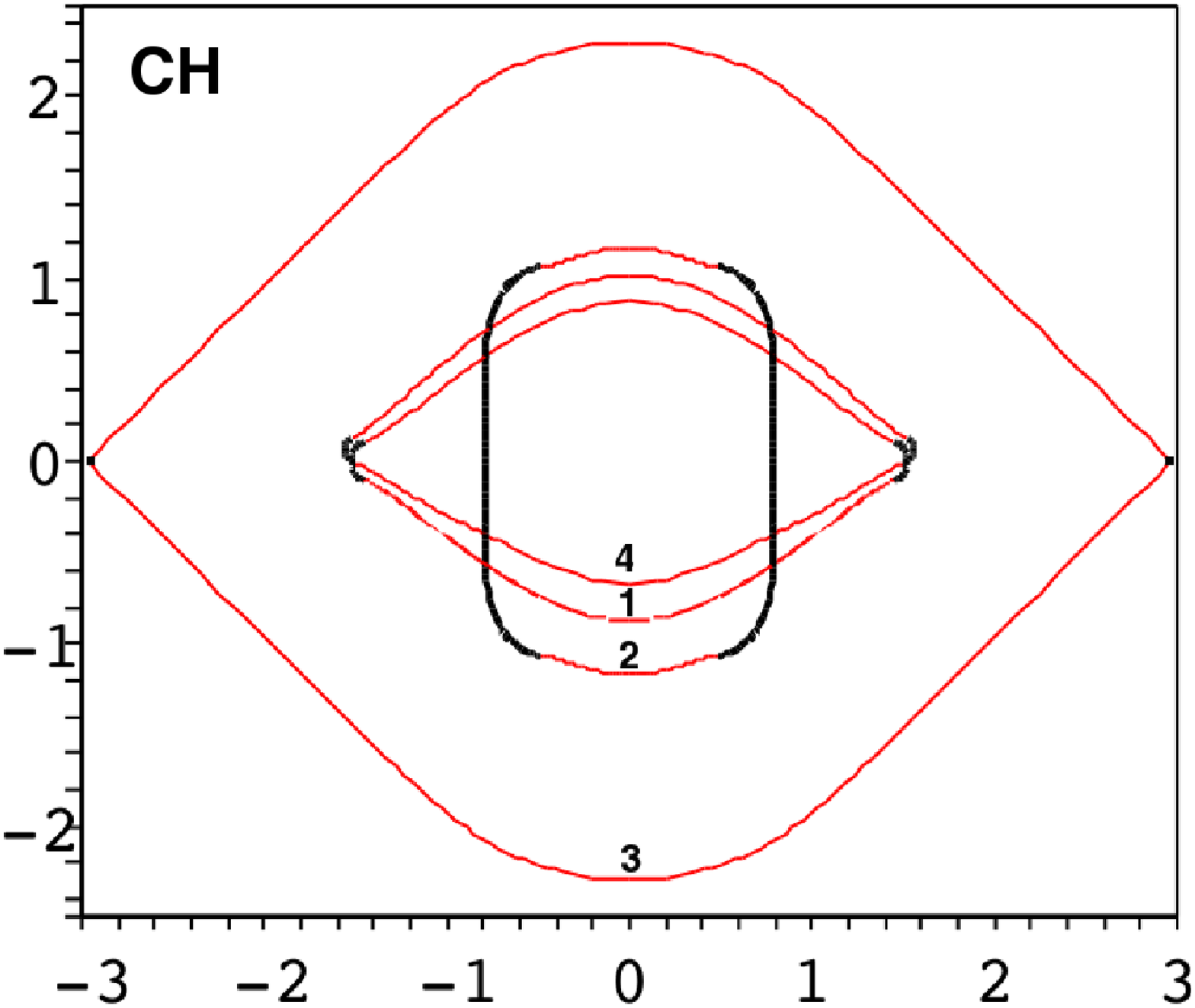}\non\\
&&\hspace{3.5cm}({\bf a})\hspace{8.0cm}({\bf b})\non\\
&&\hspace{0cm}\includegraphics[width=7.0cm]{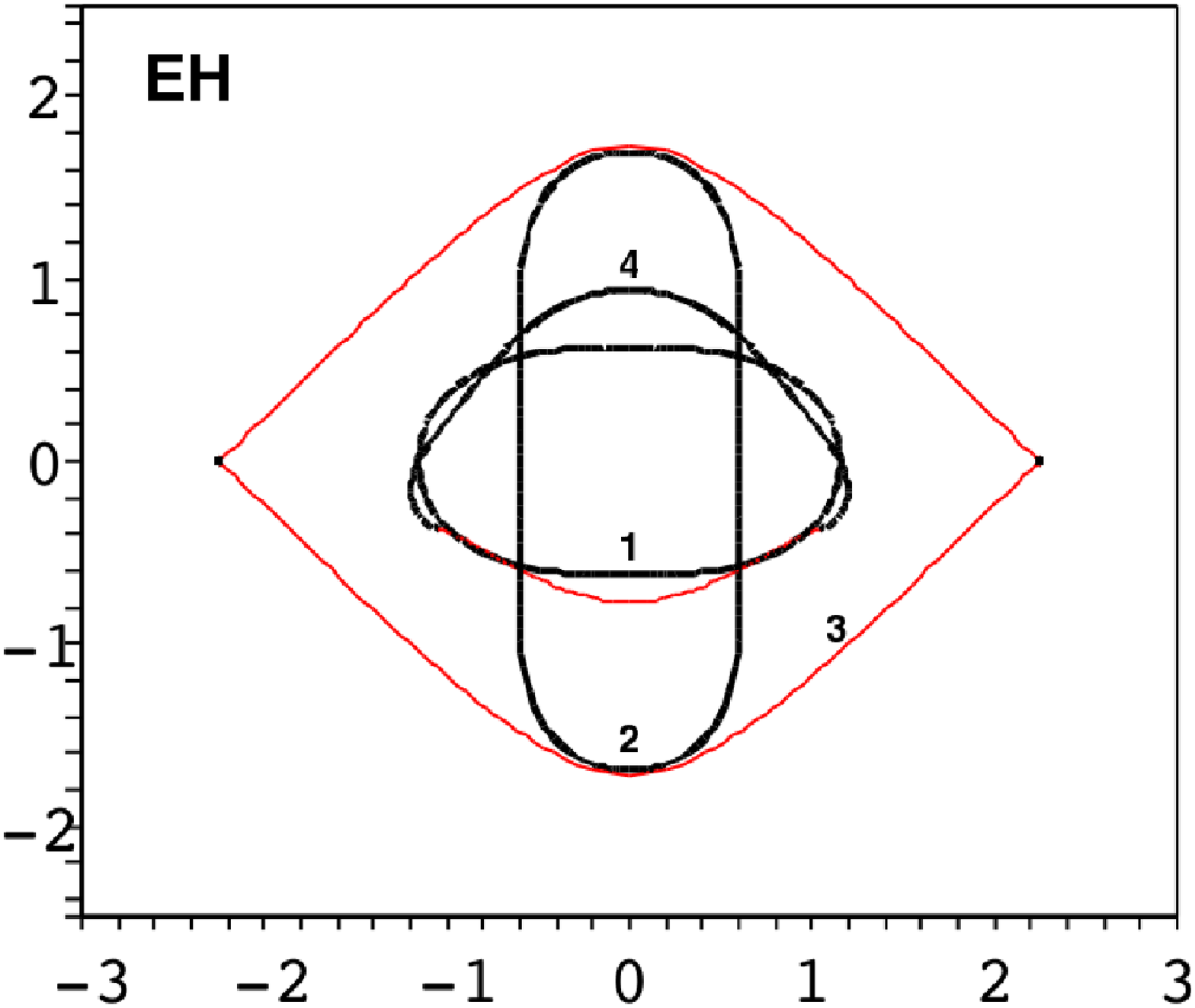}
\hspace{1.5cm}\includegraphics[width=7.0cm]{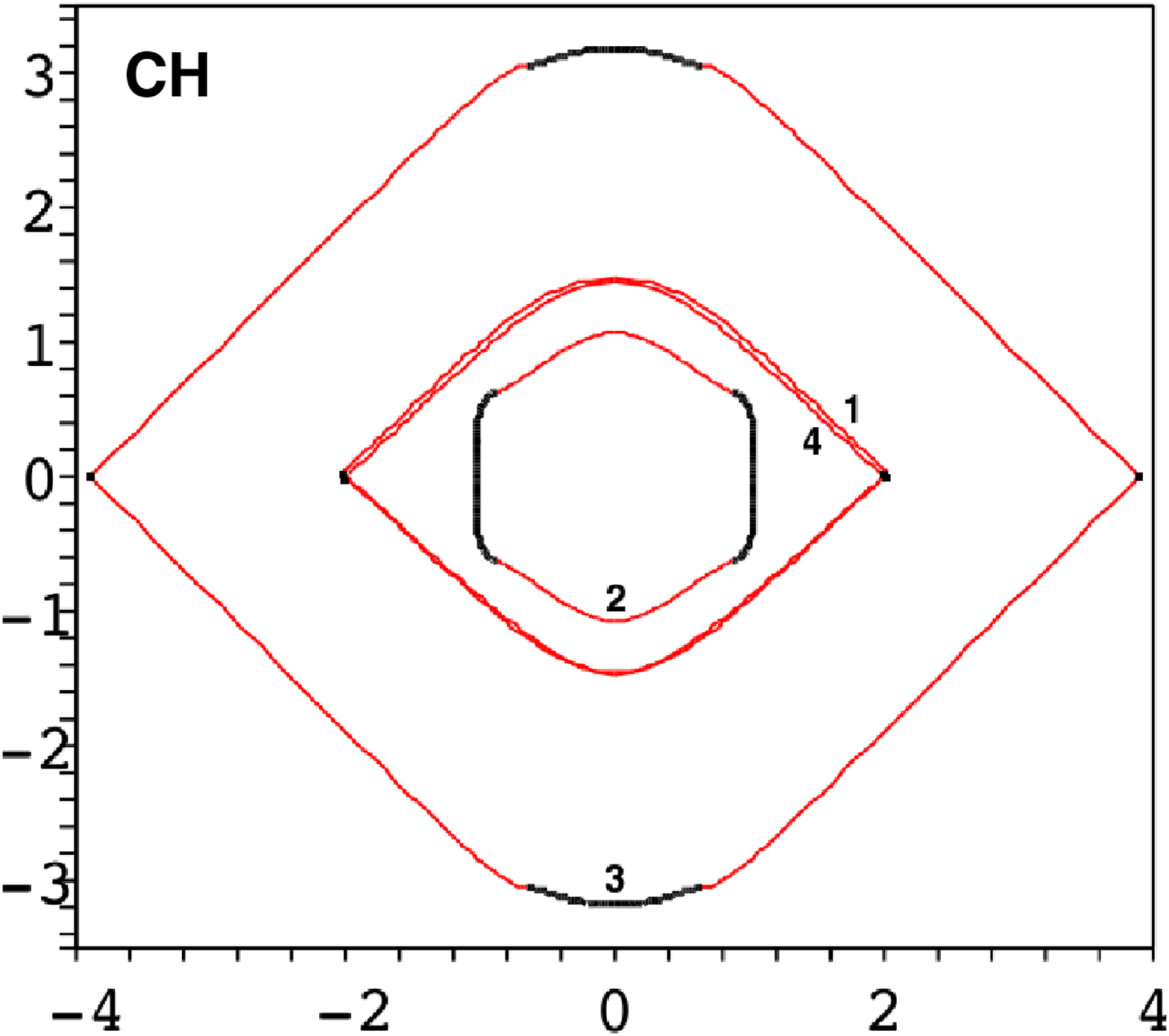}\non\\
&&\hspace{3.5cm}({\bf c})\hspace{8.0cm}({\bf d})\non\\
&&\hspace{0cm}\includegraphics[width=7.0cm]{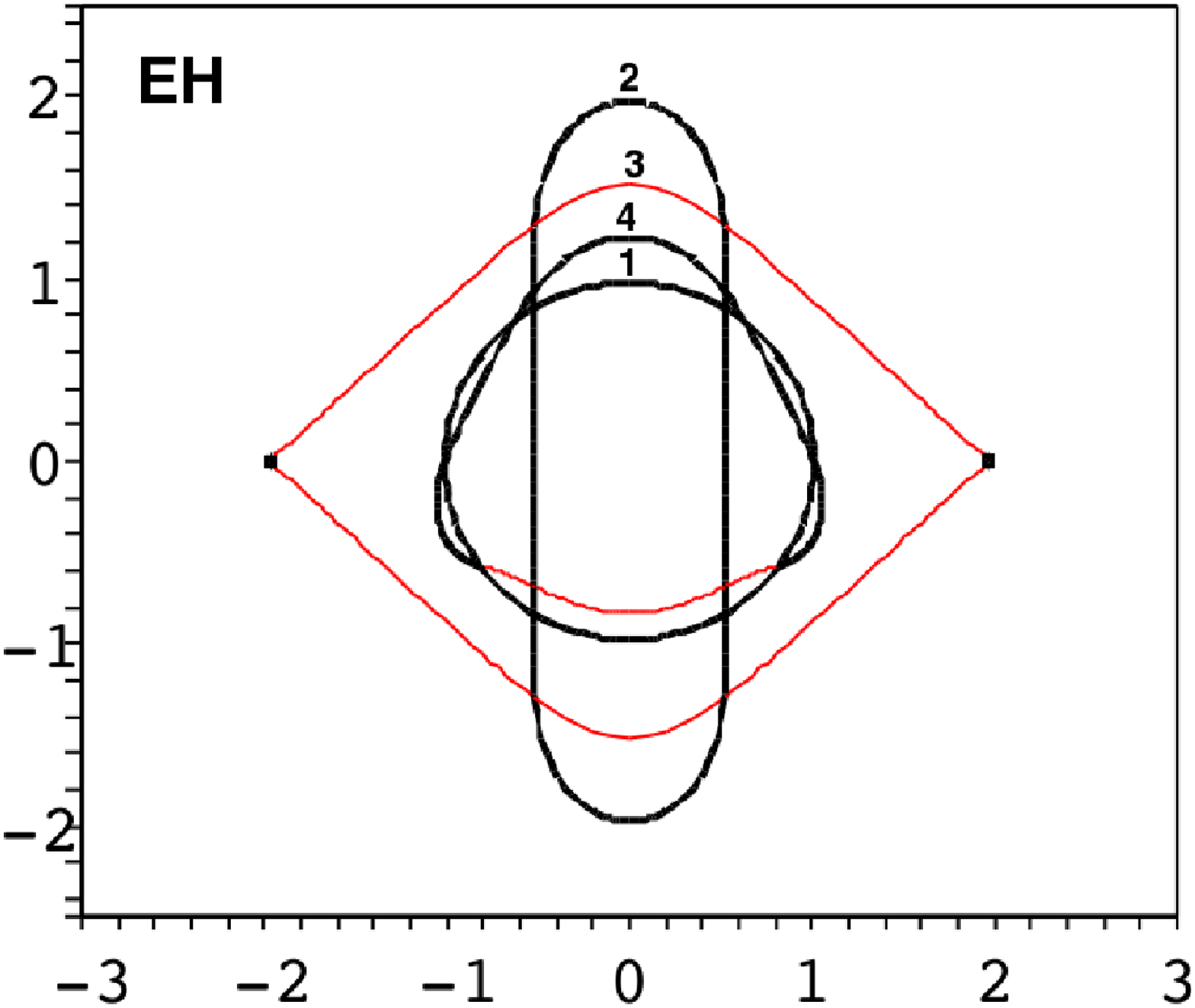}
\hspace{1.5cm}\includegraphics[width=7.0cm]{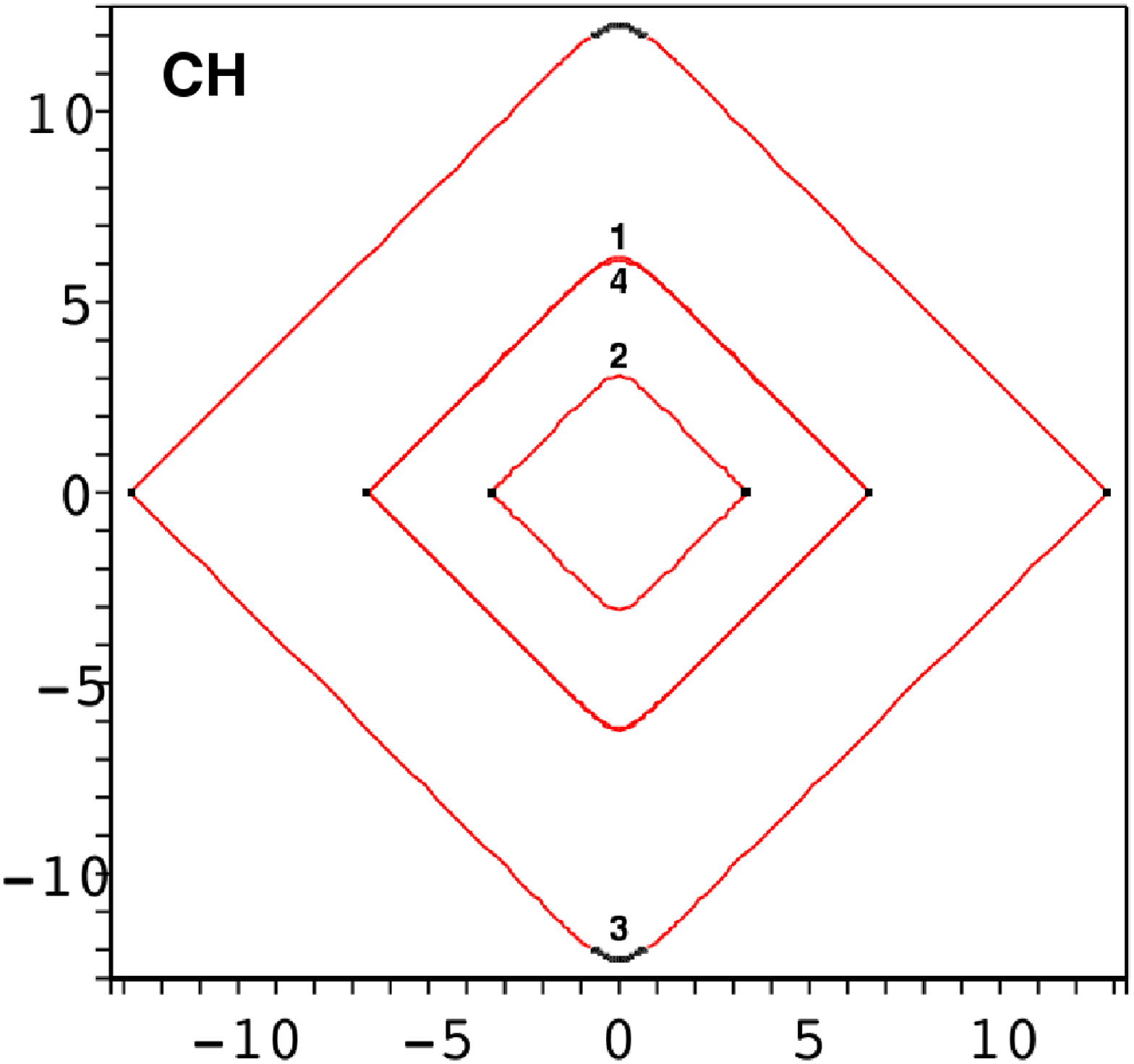}\non\\
&&\hspace{3.5cm}({\bf e})\hspace{8.0cm}({\bf f})\non
\ea
\caption{Isometric embedding of the horizon surfaces. EH stand for the outer (event) horizon, and CH stand for the inner (Cauchy) horizon. Line 1 is for undistorted horizons, line 2 is for the quadrupole distortion of $a_{2}=-2/3$, line 3 is for the quadrupole distortion of $a_{2}=2/3$, and line 4 is for the octupole distortion of $a_{3}=-1/3$. Thick (black colour) portions of the lines represent isometric embedding into a Euclidean space and thin ones (red colour) represent isometric embedding into a pseudo-Euclidean space. Plots (a) and (b) are for $a_{*}=0.99$, $\alpha\approx0.868$. Plots (c) and (d) are for $a_{*}=\sqrt{3}/2\approx0.866$, $\alpha=1/\sqrt{3}\approx0.577$. Plots (e) and (f) are for $a_{*}=0.3$, $\alpha\approx0.154$. The black dots located at $z=0$ represent small portions of the horizon surfaces embedded into Euclidean space. Due to the limited resolution, the lines there look broken, however, by the construction, all the lines are smooth. Plot (f) has almost merging lines 1 and 4. However, an isometric embedding into a Euclidean space near the equatorial plane $z=0$ exists only for undistorted inner horizon (line 1).} \label{f2} 
\end{center}
\end{figure*}

\begin{figure*}[htb]
\begin{center}
\hspace{0cm}
\ba
&&\hspace{0cm}\includegraphics[width=7.0cm]{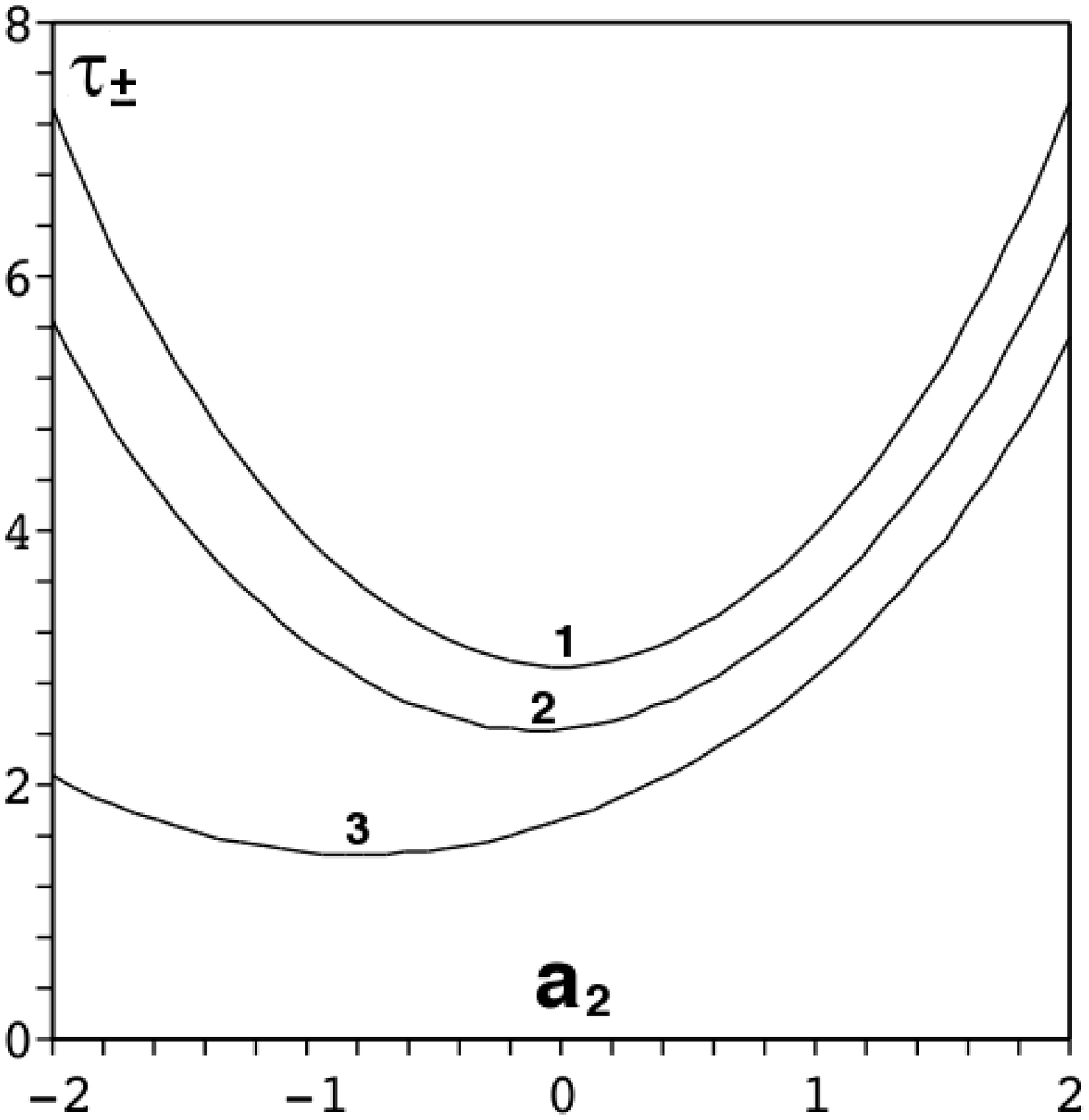}
\hspace{1.5cm}\includegraphics[width=7.0cm]{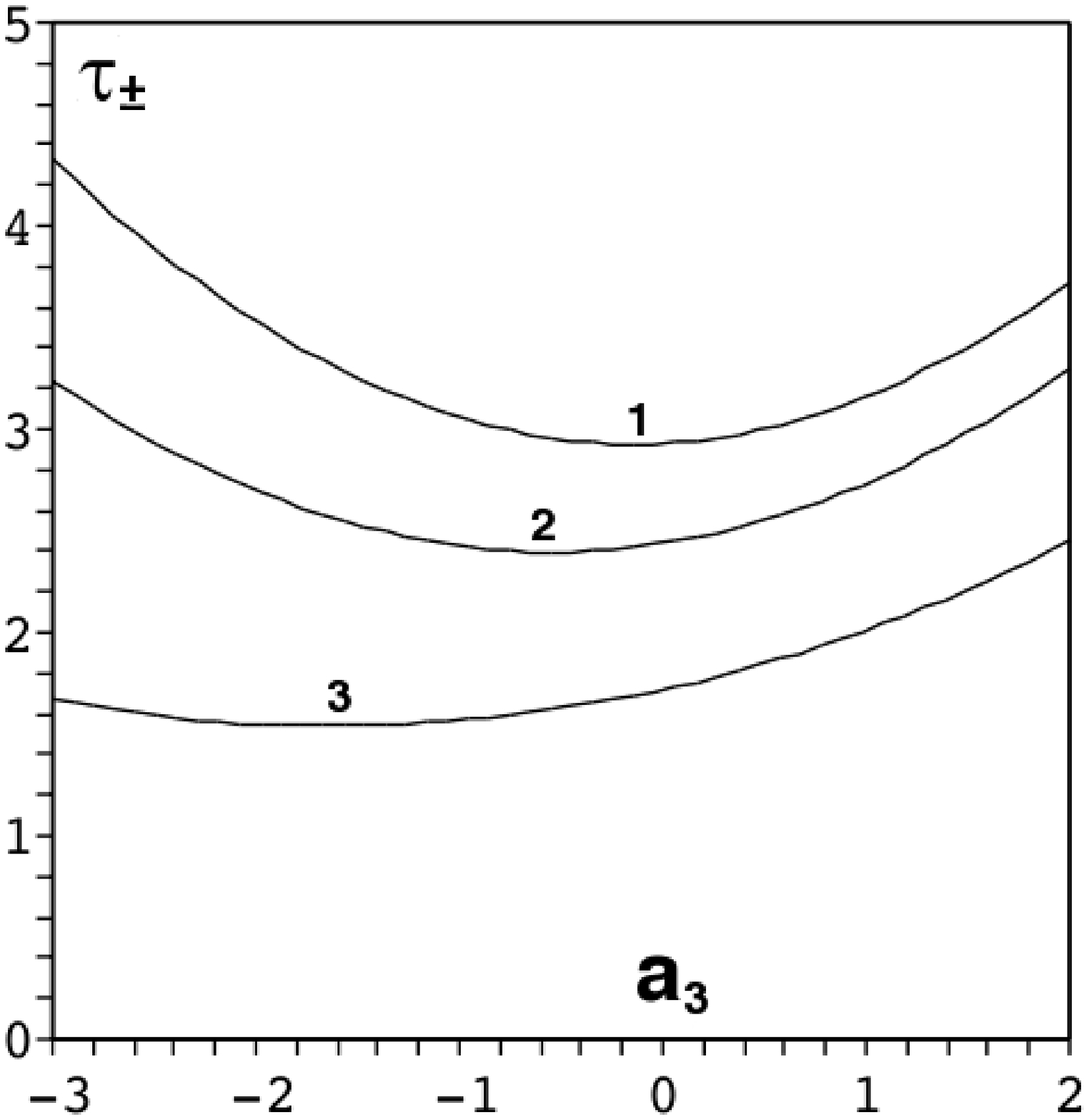}\non\\
&&\hspace{3.5cm}({\bf a})\hspace{8.0cm}({\bf b})\non
\ea
\caption{The maximal proper time of a free fall from the outer to the inner horizon along the symmetry semi-axes. Plot (a) illustrates the maximal proper time for the quadrupole distortion, which is the same for both the semi-axes, $y=\pm1$. Plot (b) illustrates the maximal proper time along the semi-axis $y=+1$ for the octupole distortion. The maximal proper time along the semi-axis $y=-1$ for the octupole distortion can be inferred from the plot (b) by the reflection $a_{3}\to-a_{3}$.  Line 1 is for $a_{*}=0.99$, $\alpha\approx0.868$, line 2 is for $a_{*}=\sqrt{3}/2\approx0.866$, $\alpha=1/\sqrt{3}\approx0.577$, and line 3 is for $a_{*}=0.3$, $\alpha\approx0.154$. } \label{f3}
\end{center}
\end{figure*}

\begin{figure*}[htb]
\begin{center}
\hspace{0cm}
\ba
&&\hspace{0cm}\includegraphics[width=7.0cm]{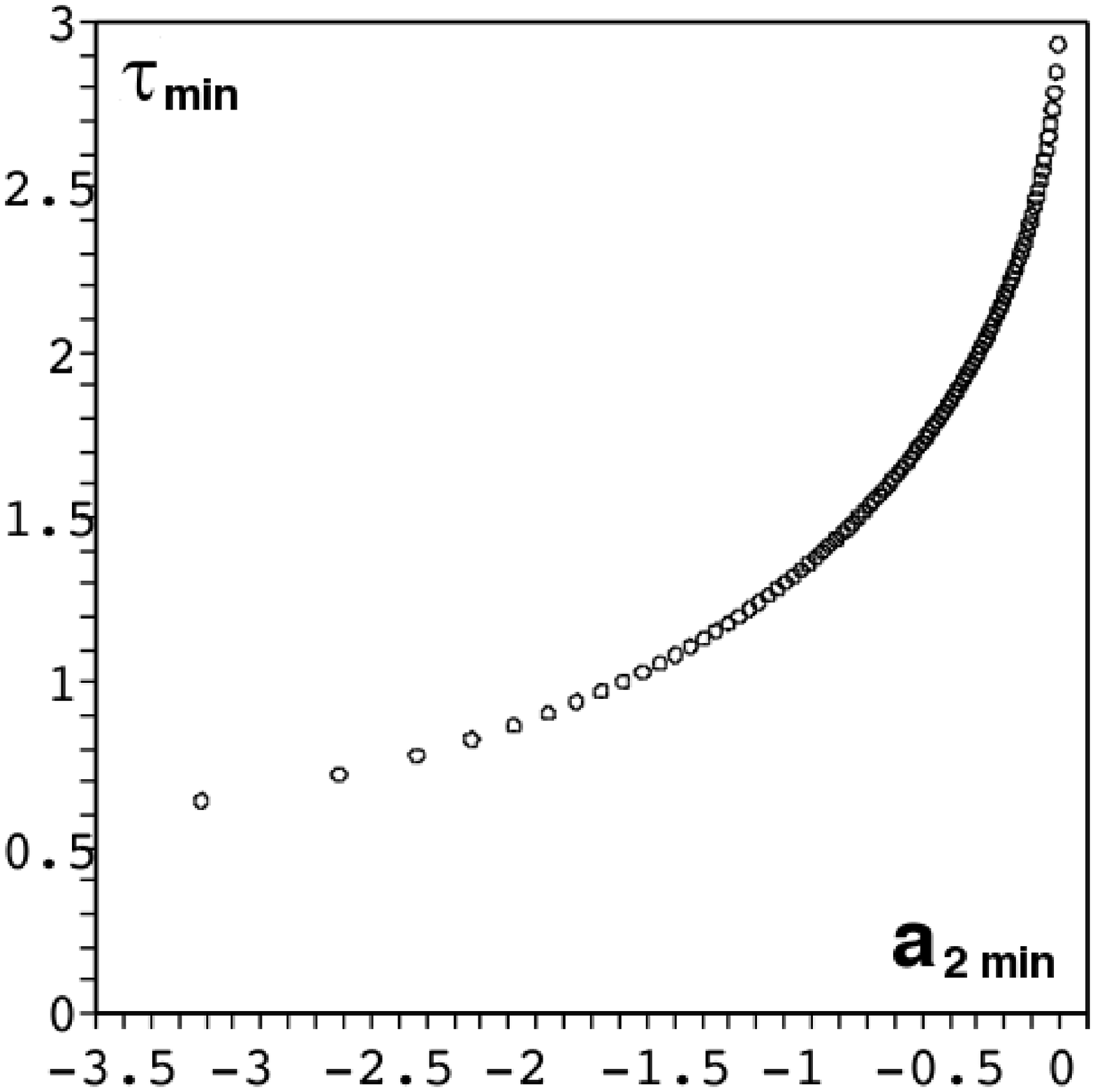}
\hspace{1.5cm}\includegraphics[width=7.0cm]{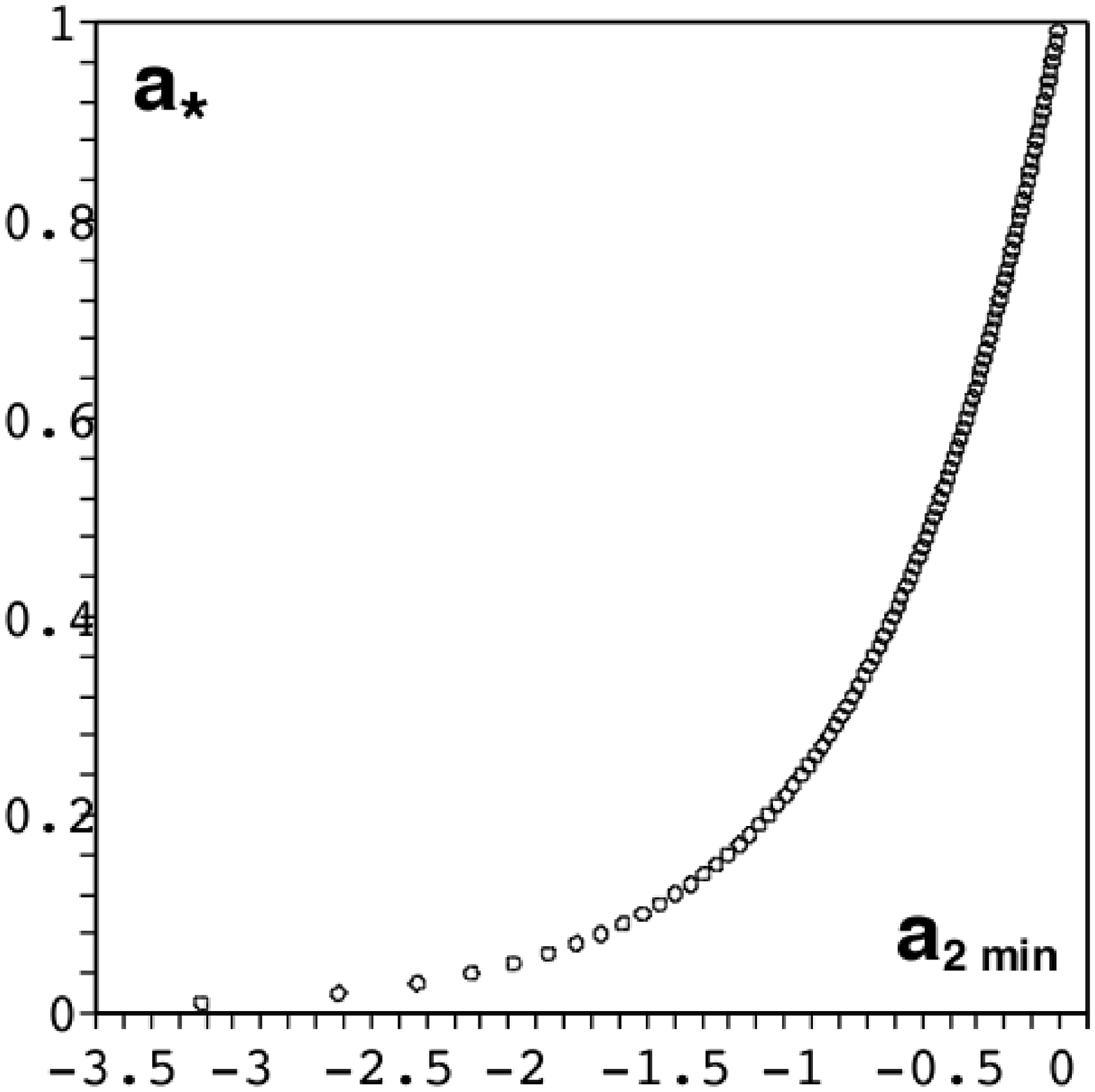}\non\\
&&\hspace{3.5cm}({\bf a})\hspace{8.0cm}({\bf b})\non\\
&&\hspace{0cm}\includegraphics[width=7.0cm]{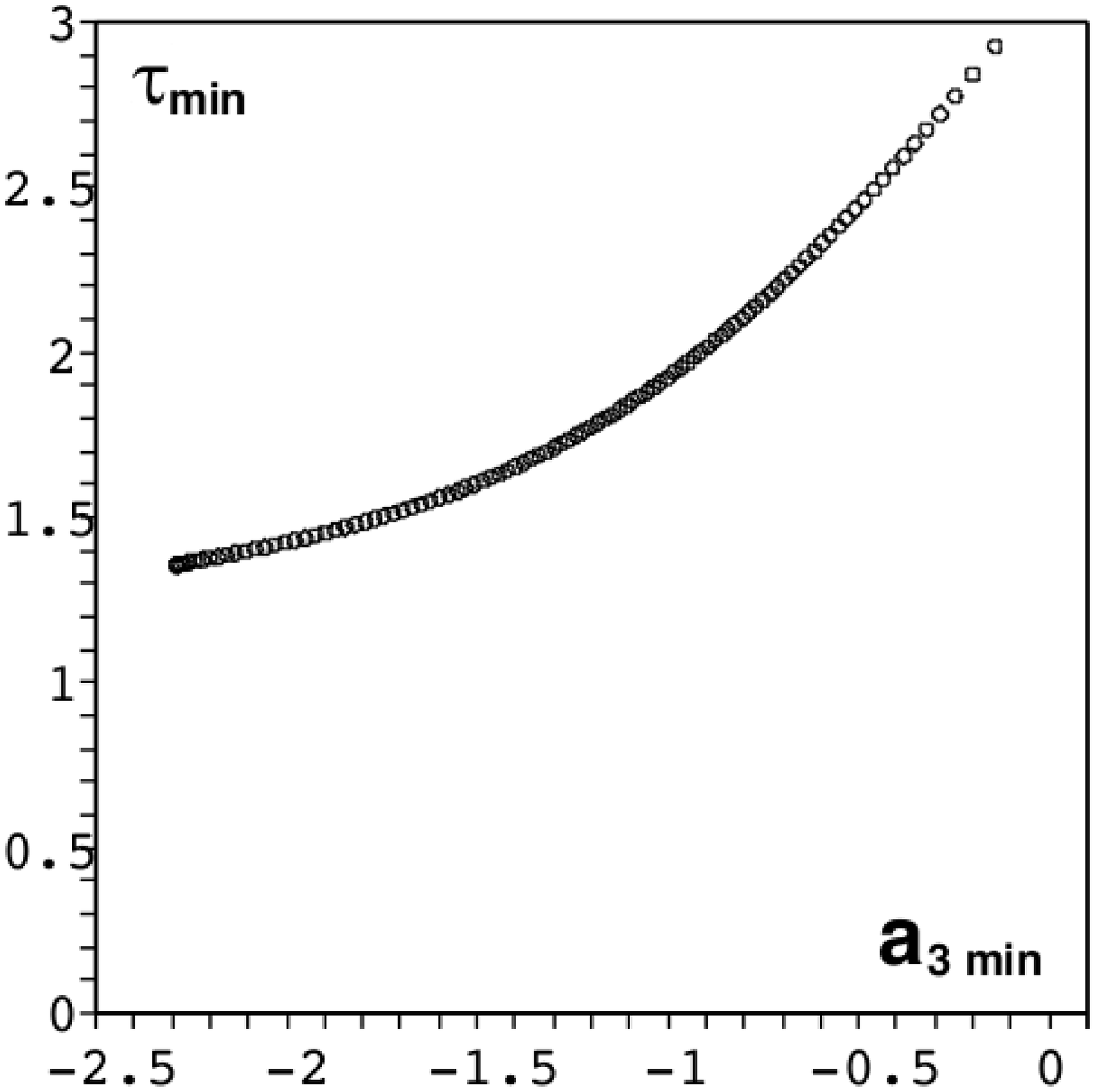}
\hspace{1.5cm}\includegraphics[width=7.0cm]{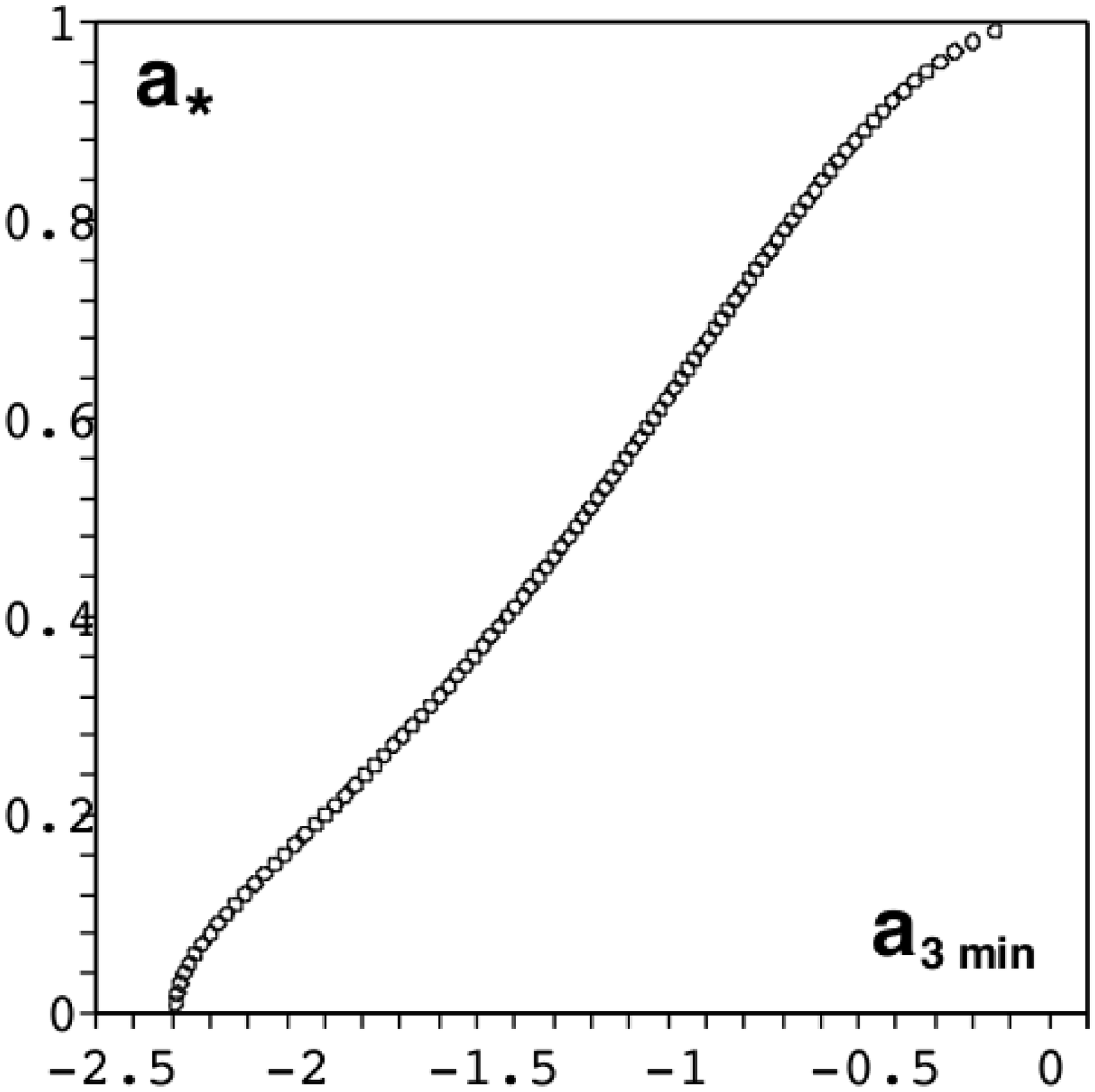}\non\\
&&\hspace{3.5cm}({\bf c})\hspace{8.0cm}({\bf d})\non
\ea
\caption{The minimal value of the maximal proper time. There are 100 points corresponding to the increment $0.01$ of $a_{*}$ generated on each of the plots.} \label{f4} 
\end{center}
\end{figure*}

\section{Geometry of Distorted Horizons}

In this Section we discuss the geometry of the distorted horizon 2-dimensional surfaces. We calculate Gaussian curvature of the horizon surfaces and present their isometric embeddings into a flat 3-dimensional space. 

\subsection{Gaussian curvature}

The 2-dimensional horizon surfaces are defined by $t=const.$ and $x=\pm 1$. The metrics on the horizon surfaces are the following:
\be\n{V.1}
d\Sigma^{2}_{\pm}=\frac{{\cal A}_{\pm}}{4\pi}\left(\frac{dy^{2}}{F_{\pm}(y)}+F_{\pm}(y)d\phi^{2}\right)\,,
\ee
where $F_{\pm}(y)$ is given by \eq{III.13}. Gaussian curvature $K_{\pm}$ is a natural measure of intrinsic curvature of a 2-dimensional surface. Gaussian curvature for the metrics \eq{V.1} reads
\be\n{V.2}
K_{\pm}=-\frac{4\pi}{{\cal A}_{\pm}}F_{\pm,yy}\,.
\ee
Here and in what follows, $(...)_{,y}=\partial(...)/\partial y$. For undistorted (Kerr) black hole Gaussian curvature of the horizon surfaces is\footnote{Note, that according to the expression \eq{IV.12b}, one has $\alpha'=\alpha$.} 
\be\n{V.3}
K'_{\pm}=\frac{8\pi}{{\cal A}'_{\pm}}\frac{(1+\alpha^{\pm2})^{2}(1-3\alpha^{\pm2}y^{2})}{(1+\alpha^{\pm2}y^{2})^{3}}\,.
\ee
It is negative at the poles $(y=\pm1)$ of the outer horizon surface for $\alpha\in(1/\sqrt{3},1)$ (see, e.g., \cite{Smarr} and \cite{Lake}), and of the inner horizon surface and for the whole the rage of $\alpha\in(0,1)$. Such surfaces cannot be isometrically embedded into a 3-dimensional Euclidean space (see, e.g., \cite{FR}).

We shall study how the distortion fields change Gaussian curvature of the horizon surfaces on the example of quadrupole and an octupole distortions. The functions $u_{\pm}(y)$ corresponding to the quadrupole and octuple distortion fields read [cf. Eq.\eq{III.3}]
\be\n{V.4}
u_{\pm}(y)=-a_2(1-y^2)\,,
\ee
and
\be\n{V.5}
u_{\pm}(y)=\mp a_3y(1-y^2)\,.
\ee
Here $a_2$ and $a_3$ are the quadrupole and the octupole moments, respectively. 

Here and in the following section, to study the horizon surfaces we consider the specific angular momentum parameter $a_{*}=a/M$ of an undistorted stationary rotating (Kerr) black hole, which, according to the expressions \eq{II.10a}, is related to the parameter $\alpha$ as follows:
\be\n{V.6}
\alpha=\frac{a_{*}}{1+\sqrt{1-a_{*}^{2}}}\,.
\ee
According to our model discussed in the previous section, $\alpha$ and so $a_{*}$ do not change when we turn on adiabatically the distortion field. This allows us to observe an effect of the distortion on the geometry of the horizon surfaces. We shall consider almost extremal value $a_{*}=0.99$, which is consistent with the measurements based on the astronomical observations giving $a_{*}>0.95$ for some black hole candidates (see \cite{NarMcC} and references therein), the value corresponding to zero Gaussian curvature at the poles of the embedded surface, $a_{*}=\sqrt{3}/2$ (see \cite{Smarr}) and the smaller value, $a_{*}=0.3$.

To illustrate the effect of the distortion we plot the Gaussian curvature of distorted and undistorted horizon surfaces,
\eq{V.2} and \eq{V.3} as a function of $y$ for a given value of $a_{*}$ and different values of $a_{2}$ and $a_{3}$ (see Fig.~\ref{f1}). Note that according to the symmetry properties of the function $u_{\pm}(y)$ [see \eq{III.3a}], the lines of $K_{\pm}$ for $a_{3}=1/3$ can be obtained from the lines corresponding to $a_{3}=-1/3$ with the aid of the reflection $y\to-y$.
One can see from the plots that for a distorted black hole the Gaussian curvature of the outer horizon surface is finite. The Gaussian curvature of the inner horizon surface can be arbitrarily large but finite for $\alpha\in(0,1)$ and it diverges for $\alpha\to0$, when the inner horizon shrinks down to a static black hole singularity. 

\subsection{Embedding}

The horizon surface is a 2-dimensional axisymmetric surface. It can be visualized by constructing its isometric embedding into a flat 3-dimensional space. To construct the embedding, we parametrize the surface as follows:
\be\n{VI.1}
\rho=\rho(y)\hh z=z(y)\,.
\ee
Let us consider the metric of a flat 3-dimensional space in the cylindrical coordinates $(z,\rho,\phi)$,
\be\n{VI.2}
dl^{2}=\epsilon dz^{2}+d\rho^{2}+\rho^{2}d\phi^{2}\,.
\ee
Here $\epsilon=-1$ corresponds to a pseudo-Euclidean space and $\epsilon=1$ corresponds to a Euclidean one. We shall consider an isometric embedding of the surface into a pseudo-Euclidean space, if such a surface cannot be isometrically embedded into a Euclidean one.\footnote{As it was already mentioned in the previous subsection, a 2-dimensional axisymmetric surface cannot be isometrically embedded into a 3-dimensional Euclidean space if its Gaussian curvature is negative in the vicinity of the symmetry axis (see, e.g., \cite{FR}). The reverse is not generally true. A failure to construct such an embedding does not necessarily imply that the Gaussian curvature is negative.} Using the expressions \eq{VI.1} and \eq{VI.2} we have the surface metric induced by the embedding geometry 
\be\n{VI.3}
dl^{2}=(\epsilon z_{,y}^{2}+\rho_{,y}^{2})dy^{2}+\rho^{2}d\phi^{2}\,.
\ee
Let us consider dimensionless metrics of the horizon surfaces $d\bar{\Sigma}_{\pm}^{2}$ defined as follows [cf. Eq.\eq{V.1}]:
\be\n{VI.4}
d\Sigma_{\pm}^{2}=\frac{{\cal A}_{\pm}}{4\pi}d\bar{\Sigma}_{\pm}^{2}\hh d\bar{\Sigma}_{\pm}^{2}=\frac{dy^{2}}{F_{\pm}}+F_{\pm}d\phi^{2}\,.
\ee
Matching the metrics \eq{VI.3} and \eq{VI.4} we derive the embedding map
\be
\rho_{\pm}=\sqrt{F_{\pm}}\hh z_{\pm}=\int^{y}_{0}{\cal Z}_{\pm}dy\hh {\cal Z}_{\pm}^{2}=\frac{\epsilon}{4F_{\pm}}(4-F_{\pm,y}^{2})\,.\n{VI.5}
\ee
According to this map, for $|F_{\pm,y}|>2$ an isometric embedding of the horizons surface into a 3-dimensional Euclidean space ($\epsilon=1$) is not possible, and we have to take $\epsilon=-1$.  

Because of the space-time axial symmetry, to illustrate the embeddings it is enough to consider a line in the $(z,\rho)$-plane given by the embedding map \eq{VI.5}. Then, the embedding is represented by a 2-dimensional surface generated by a revolution of the line about the vertical $z$-axis, which passes through the centre $(0,0)$ of each plot. Isometric embeddings of the horizon surfaces are illustrated in Fig.~\ref{f2}. Note that according to the symmetry properties of the function $u_{\pm}(y)$ [see \eq{III.3a}], embeddings for $a_{3}=1/3$ can be obtained by the reflection of the lines corresponding to $a_{3}=-1/3$ with respect to the horizontal axis $z=0$. One can see that for the undistorted case, for the quadrupole distortion of negative quadrupole moment $a_{2}$, and for the octupole distortion the outer horizon surface has bigger relative portion which is embeddable into Euclidean space than that of the inner horizon surface. The situation is opposite in the case of the quadrupole distortion of positive $a_{2}$. 

\section{Free fall from the outer to the inner horizon}

Let us now consider how the distortion changes the black hole interior region between its horizons. We shall be interested whether the horizons can come closer to each other or move away due to the distortion. To do it, we shall calculate the maximal proper time of a test particle dropped from the outer horizon surface and freely falling to the inner horizon along the symmetry semi-axes $y=\pm1$. This maximal proper time corresponds to a timelike geodesics of zero energy and azimuthal angular momentum, which are defined as follows:
\be\n{VII.1}
E=-u^{\alpha}\xi^{\beta}_{(t)}g_{\alpha\beta}\hh L_{\phi}=u^{\alpha}\xi^{\beta}_{(\phi)}g_{\alpha\beta}\,,
\ee
where $u^{\alpha}$ is 4-velocity of the particle. This implies that the coordinates $(t,y,\phi)$ remain constant along the particle's geodesic, and the coordinate $x$ changes from $1$ to $-1$ during the fall. Substituting $y=\pm1$ into the metric functions we derive the following useful relations:
\ba
f_{\pm}&=&\alpha\,e^{\pm2u_{\pm}(x)}\hh h_{\pm}=-\alpha^{2}/f_{\pm}\,,\n{VII.2a}\\
U_{\pm}&=&u_{\pm}(x)+u_{0}\hh V_{\pm}=0\,,\n{VII.2b}
\ea
where $u_{\pm}(x)$ is given by \eq{III.3}, where $y$ should be replaced with $x$. The upper sign stands for the upper semi-axis $y=+1$, what corresponds to the fall from the black hole's ``north pole'' and the lower sign stands for the lower semi-axis $y=-1$, what corresponds to the fall from the black hole's ``south pole''. Using the metric \eq{II.1} we derive the expression for the  proper time in the units of the outer horizon surface radius $R_{+}=\sqrt{{\cal A}_{+}/(4\pi)}$,
\ba
\tau_{\pm}&=&\frac{1}{2(1+\alpha^{2})^{1/2}}\int_{-1}^{1}\frac{{\cal T}_{\pm}dx}{(1-x^{2})^{1/2}}\,,\n{VII.3}\\
{\cal T}_{\pm}&=&\left[(1+x+\alpha^{2}[1-x])^{2}e^{-2u_{\pm}(x)}+4\alpha^{2}e^{2u_{\pm}(x)}\right]^{1/2}\,.\non
\ea
Let us remind the reader that according to our model, the distortion field changes adiabatically, so the horizon surface area ${\cal A}_{+}$, and so $R_{+}$, remain constant and equal to those of undistorted (Kerr) black hole. We illustrate in Fig.~\ref{f3} how the proper time depends on the distortion field, in particular on the quadrupole $a_{2}$ and the octupole $a_{3}$ moments, for different values of the specific angular momentum parameter $a_{*}$. The behaviour of the maximal proper time is generic. Namely, for a given $a_{*}\in(0,1)$ there is a characteristic minimum $\tau_{\pm\text{min}}$ corresponding to $a_{2\text{min}}$ and $a_{3\text{min}}$ which indicates the closes approach of the horizons due to the distortion field. One can calculate numerically how the minimal proper time $\tau_{\text{min}}$ depends on the corresponding ``minimal'' values of the octupole $a_{2\text{min}}$ and the quadrupole $a_{3\text{min}}$ moments and find the corresponding values of the specific angular momentum parameter $a_{*}$. The results are presented in Fig.~\ref{f4}. We see that both $\tau_{\text{min}}$ and  $a_{*}$ are monotonically increasing with the increase of $a_{2\text{min}}$ and $a_{3\text{min}}$. As a summary, one can say that the minimal value of the maximal proper time is largely controlled by the specific angular momentum parameter $a_{*}$ and a distortion field can bring the horizons close to each other but they do not merge. This situation is different for a free fall from the outer horizon to the singularity of a static (Schwarzschild) black hole distorted by a quadrupole field, when $\tau_{\text{min}}$ can be arbitrarily small for an arbitrary large negative quadrupole moment \cite{FS}. But the situation is qualitatively the same for an octupole distortion of a static black hole where is a minimal finite value of $\tau$. 

\section{The space-time curvature invariants and singularities}

In this section we discuss some general properties of the space-time \eq{II.1}. We establish a discrete symmetry of the space-time from which the duality transformation between the black hole horizons directly follows. Such a symmetry is an inherent property of the metric and lies in its very construction. This discrete symmetry allows to relate the space-time curvature invariants calculated on the black hole horizons. We also analyze the space-time curvature singularities and discuss their location. 

\subsection{The space-time discrete symmetry and curvature invariants} 

A closer look at the duality transformation between the black hole horizons \eq{III.16} suggests that it may originate from some general property of the space-time.  Indeed, the metric \eq{II.1} reveals that it has a discrete symmetry,
\be\n{VII.1}
t\to-t\hh x\to-x\hh y\to-y\hh \alpha\to \alpha^{-1}\,.
\ee
On the horizon surfaces, this symmetry reduces to the duality transformation \eq{III.16}. The symmetry implies that a space-time curvature invariant $I(x,y,\alpha,m,a_{n})$ has the following symmetry property:
\be\n{VII.2}
I(x,y,\alpha,m,a_{n})=I(-x,-y,\alpha^{-1},m,a_{n})\,.
\ee
Let us define the values of the invariant on the black hole horizons as follows:
\be\n{VII.3}
I_{\pm}(y,\alpha,m,a_{n})=I(x=\pm1,y,\alpha,m,a_{n})\,.
\ee
Then, the symmetry property \eq{VII.2} implies that on the horizons we have the following duality relation:
\be\n{VII.4}
I_{\pm}(y,\alpha,m,a_{n})=I_{\mp}(-y,\alpha^{-1},m,a_{n})\,.
\ee
This duality relation implies that if the metric is regular on the outer horizon for all $y\in[-1,1]$ and $\alpha$ replaced with $\alpha^{-1}$, then it is regular on the inner (Cauchy) horizon, and vice versa. 

The location of the space-time curvature singularities is defined by $B=0$ (see \cite{Tom,Man}). Indeed, the Kretschmann and the Chern-Pontryagin invariants, as well as the metric determinant diverge at $B=0$. According to the expression for the metric function $B(x,y)$ [see \eq{II.2b}], we have $B\ne0$ at the outer horizon for whole the range of $y\in[-1,1]$ and $\alpha\in(0,1)$, as well as for $\alpha$ replaced with $\alpha^{-1}$, and vice versa. Thus, the space-time is regular at the distorted black hole horizons.   

\subsection{Singularities}

Let us now to find the location of the space-time singularities. To begin with let us present the metric functions $f$ and $h$ [see Eqs.\eq{II.3a} and \eq{II.3b}] in the following convenient form:
\be\n{VII.5}
f=\alpha e^{p}\hh h=-\alpha e^{q}\hh fh=-\alpha^{2}e^{-2V_1}\,,
\ee
where $p$ and $q$ are functions of $x$ and $y$, and $V_1$ is the linear in $a_{n}$'s part of $V$ [see Eq.\eq{II.4b}]. Then, the equation for the space-time singularities $B=0$ [see Eq.\eq{II.2b}] is equivalent to the following set of equations:
\be\n{VII.6}
x=-\frac{1+\alpha^{2}e^{-2V_{1}}}{1-\alpha^{2}e^{-2V_{1}}}\hh y=-\frac{e^{p}-e^{q}}{e^{p}+e^{q}}\,.
\ee
For $\alpha\in(0,1)$ the first equation implies that $|x|>1$, what means that the space-time singularities are located behind the inner horizon and, if any, outside of the outer horizon, i.e., there are no space-time singularities located in between the horizons. The second equation implies that $0\leq|y|<1$, what means that there are no space-time singularities on the symmetry axis. Let us now formulate a {\em sufficient condition} that there are no singularities located outside the outer horizon, i.e., in the region $x>1$. Using positivity of the exponential function and the first equation we derive the {\em sufficient condition}
\be\n{VII.7}
V_{1}|_{x>1}\geq0\,.
\ee
Let us note that the sufficient condition $a_{2n-1}=0$, $a_{2n}\leq0$, $n=1,2,...$ for the absence of the singularities in the region outside the outer horizon proposed in \cite{Man,ManQ} is not valid. To illustrate this, let us consider a hexadecapole distortion field defined by the multipole moments $(a_{0}, a_{1}, a_{2}, a_{3}=-a_{1}, a_{4})$. For this field the function $V_{1}$ takes the from
\ba\n{VII.8}
V_{1}&=&\{a_{4}x(x^{2}[1-5y^{2}]-3[1-y^{2}])\non\\ 
&-&a_{3}y[3x^{2}-1]-2a_{2}x\}(1-y^{2})\,.
\ea
Let us now take $a_{3}=0$ and consider the second equation in the system \eq{VII.6}. Using the properties of the Legendre polynomials \eq{II.5c} and the expressions \eq{II.3a} and \eq{II.3b} one can see that $y=0$ is a solution of the second equation in the system \eq{VII.6}. Note that $y=0$ is a solution for any type of distortion defined by even multipole moments. Substituting this solution into the first equation in the system \eq{VII.6} we derive
\be\n{VII.9}
a_{4}=\frac{2a_{2}}{x^{2}-3}+\frac{1}{2x(x^{2}-3)}\ln\left(\frac{\alpha^{2}[x-1]}{[x+1]}\right)\,.
\ee
Taking, for example, $x=2$, $\alpha=1/\sqrt{3}$, $a_{2}=-1$, we find $a_{4}=-2-\ln(3)/2$. Thus, for these negative values of $a_{2}$, $a_{4}$ there is a singularity located at the intersection of the lines $x=2$ and $y=0$ in the $(x,y)$ plane. This is a counterexample to the sufficient condition proposed in \cite{Man,ManQ}. However, as it was illustrated in \cite{Man,ManQ}, for a pure quadrupole distortion defined by $a_{2}<0$ there are no singularities located outside of the outer horizon. This can be confirmed by the {\em sufficient condition} \eq{VII.7} for $V_{1}$ given in \eq{VII.8} with $a_{3}=a_{4}=0$.

The example of the hexadecapole distortion illustrates that it is rather impossible to satisfy the {\em sufficient condition} for a general type of distortion. Thus, in general, there are space-time curvature singularities outside the black hole outer horizon. They are represented by a discrete set of points in the $(x,y)$ plane, which are solutions to the system \eq{VII.6}. These points represent rings in the space-time around the black hole. The singular points lie on the curves defined by the equation $A=0$. Indeed, using the expression \eq{II.2a} one can see that solutions of the system \eq{VII.6} are as well solutions of the equation $A=0$. Note that equation $A=0$ defines stationary limit surfaces. These surfaces are represented by lines in the $(x,y)$ plane. On the other side, the distortion field is static. Thus, one may assume that the singularities outside the outer horizon originate from the static nature of the distortion field which is ``incompatible'' with a stationary limit surface, where nothing can be hold at rest. However, such singularities are absent for a quadrupole distortion field of $a_{2}<0$, for which the {\em sufficient condition} \eq{VII.7} is satisfied. Thus, the situation is not trivial and an additional analysis of the origin of these singularities, which is beyond of scope of this paper, is necessary. Using these results, one can say that the {\em local stationary rotating distorted black hole} solution should be defined in the region near the outer horizon which does not include these singularities.

\section{Conclusion}
 
The main goal of this work was to study how a static and axisymmetric distortion field affects a stationary rotating black hole horizons and interior region. It was found that there is a duality transformation [see Eqs.\eq{III.15} and \eq{III.16}] between the inner and outer horizons of the black hole. This duality transformation differs from the duality transformation between the inner and outer horizons of a distorted, static, and electrically charged black hole \cite{AFS} and from the duality transformation between the horizon and stretched singularity of a distorted, static, vacuum black hole \cite{FS}. It turns out that the duality transformation is directly related to the discrete symmetry [see Eq.\eq{VII.1}] of the space-time. This symmetry implies a relation between space-time curvature invariants calculated on the black hole horizons. We constructed Smarr's formulas and formulated the laws of thermodynamics for both the horizons. The constructed Smarr's formulas and the laws of thermodynamics are related by the duality transformation as well. 

To analyze the effect of distortion field on a stationary rotating black hole we restricted ourselves to two types of the distortion, a quadrupole and an octupole ones, defined by a quadrupole and octupole moments, respectively. It was found that such distortion fields strongly affect the Gaussian curvature of the horizon surfaces, which takes both positive and negative values. The curvature is finite on both the horizons. The maximal absolute values of the Gaussian curvature of the inner horizon surface are greater than those of the outer horizon surface. The curvature can become arbitrarily large but finite for an arbitrarily small but non-vanishing specific angular momentum parameter. This is quite expected, because for very small values of the parameter, the inner horizon comes closer to the black hole singularity.  To illustrate the effect of the distortion on the black hole horizons geometry we constructed isometric embeddings of the horizon surfaces into a flat 3-dimensional space. The embeddings show that the inner horizon has larger portions which cannot be isometrically embedded into a 3-dimensional Euclidean space than the outer one.  The regions not isometrically embeddable into a 3-dimensional Euclidean space grow with the increase of the specific angular momentum parameter and, generally, with the increase of the distortion field. An analysis of the maximal proper time of a free fall from the outer to the inner horizon along the symmetry semi-axes showed that the distortion field noticeably affects the black hole interior. There is some minimal non-zero value of the quadrupole and octupole moments when the time becomes minimal. It indicates the closest approach of the horizons due to the distortion. The minimal value of the maximal proper time monotonically increases with an increase of the specific angular momentum parameter for both types of the distortion. Finally, we analyzed the space-time curvature singularities and formulated the {\em sufficient condition} for their absence outside the black hole outer horizon. We found that the region between the black hole horizons is regular, while the region behind the inner horizon and outside the outer one generally has curvature singularities. These singularities represent rings around the symmetry axis. The singularities behind the inner horizon are natural, while the singularities outside the outer horizon are generally unavoidable. They appear on a stationary limit surface. This result suggests that the {\em local stationary rotating distorted black hole} solution should be considered up to the region containing these singularities.

The duality transformation and the discrete symmetry of the space-time seem to be an inherent property of the Weyl-type solution. The corresponding Einstein equations are equivalent to two complex Ernst equations which can be considered as an integrability condition for an associated linear problem (see, e.g., \cite{AH2,AH3}). The linear problem can be formulated by means of a Lax pair construction for four- and  higher-dimensional models of gravity (see, e.g., \cite{Gal,GR}). A Lax pair is directly related to a generation of an infinite number of soliton-type solutions, starting from a known one. This lead to a complete integrability of the dynamical system (see, e.g., \cite{Bel1,Bel2}). Thus, one may expect that a certain duality transformation may exist between horizons of some solutions of these models, which possess a certain group of isometries. Recently, Abdolrahimi, Kunz, and Nedkova constructed a new exact solution within the Weyl formalism which represent a 5-dimensional Myers-Perry black hole with a single angular momentum in an external gravitational field \cite{AKN}. It is a higher-dimensional generalization of the 4-dimensional distorted stationary rotating black hole studied in this work. The solution has many interesting features related to its inherent symmetries. Some of them were already explored in \cite{AKN}. Moreover, as it was recently showed by V. P. Frolov and A. V. Frolov \cite{FF}, the Weyl-type solutions admit a construction of the so-called hybrid black holes obtained by gluing the external Kerr space-time to the internal static Weyl metric, which represents distorted axisymmetric, static, vacuum black hole. The multipole moments of the distortion field are defined by the Kerr black hole's mass and angular momentum. It would be interesting as well to investigate a relation between the black hole horizons as isolated horizons, the future outer trapping horizons, and marginally trapped surfaces, as it was done for a distorted static (Schwarzschild) black hole by Pilkington, Melanson, Fitzgerald, and Booth in \cite{PMFB}. Apparently, the rich structure of the Weyl-type solutions is far to be completely explored.

\begin{acknowledgments}

This research was supported  by the Natural Sciences and Engineering
Research Council of Canada. The author is thankful to Prof. Don N. Page for bringing to his attention the work \cite{MarOri} and many useful discussions and comments.

\end{acknowledgments}


\end{document}